\documentclass[a4paper]{article}
\usepackage{RR}
\usepackage{hyperref}
\RRdate{Octobre 2006}
\RRauthor{
Mohab Safey El Din\thanks{{\tt Mohab.Safey@lip6.fr}}%
  \and
Philippe Tr\'ebuchet\thanks{{\tt Philippe.Trebuchet@lip6.fr}}%
}
\authorhead{Safey El Din \& Tr\'ebuchet}
\RRtitle{Th\'eor\`eme de B\'ezout bi-homog\`ene fort et applications en
  g\'eom\'etrie alg\'ebrique r\'eelle effective}
\RRetitle{Strong bi-homogeneous B\'ezout theorem and its use in effective real
  algebraic geometry}
\titlehead{Bi-homogeneous B\'ezout theorem and its use in real geometry}
\RRresume{Soit $(f_1, \ldots, f_s)$ une famille de polyn\^omes dans $\Q[X_1,
\ldots, X_n]$ (o\`u $s\leqslant n-1$) de degr\'e born\'e par $D$. On suppose
que $\<f_1, \ldots, f_s\>$ engendre un id\'eal radical, et d\'efinit une
vari\'et\'e alg\'ebrique lisse $\mathcal{V}\subset\C^n$. Consid\'erons une
projection $\pi:\C^n\rightarrow\C$. On prouve que le degr\'e du lieu critique
de $\pi$ restreinte \`a $\mathcal{V}$ est born\'e par
$D^s(D-1)^{n-s}\binom{n}{n-s}$. Ce r\'esultat est obtenu en deux temps. Tout
d'abord, on caract\'erise les points critiques de $\pi$ restreinte \`a
$\mathcal{V}$ comme projections des solutions du syst\`eme de Lagrange pour
lequel on exhibe une structure bi-homog\`ene. Puis, on prouve une th\'eor\`eme
de B\'ezout bi-homog\`ene, qui borne la somme des degr\'es des composantes
\'equi-dimensionnelles du radical d'un id\'eal engendr\'e par une famille de
polyn\^omes bi-homog\`enes. Ce r\'esultat est am\'elior\'e dans le cas o\`u
$(f_1, \ldots, f_s)$ est une suite r\'eguli\`ere. De plus, on utilise la
formulation Lagrangienne pour d\'ecrire un algorithme calculant au moins un
point par composante connexe d'une vari\'et\'e alg\'ebrique r\'eelle
lisse. Cet algorithme g\'en\'eralise au cas non \'equi-dimensionnel celui de
Safey El Din et Schost. L'estimation de la taille de la sortie de notre
algorithme donne de nouvelles (et meilleures) bornes sur le premier nombre de
Betti d'une vari\'et\'e alg\'ebrique r\'eelle lisse. Finalement, on montre
qu'une instance probabiliste de notre algorithme est de complexit\'e
polynomiale en $n$, $s$, $D^s(D-1)^{n-s}{\binom{n}{n-s}}$ et la complexit\'e
d'\'evaluation de $f_1, \ldots, f_s$.}

\RRabstract{Let $(f_1, \ldots, f_s)$ be a polynomial family in $\Q[X_1,
\ldots, X_n]$ (with $s\leqslant n-1$) of degree bounded by $D$. Suppose that
$\<f_1, \ldots, f_s\>$ generates a radical ideal, and defines a smooth
algebraic variety $\mathcal{V}\subset\C^n$. Consider a projection
$\pi:\C^n\rightarrow\C$. We prove that the degree of the critical locus of
$\pi$ restricted to $\mathcal{V}$ is bounded by
$D^s(D-1)^{n-s}\binom{n}{n-s}$. This result is obtained in two steps.  First
the critical points of $\pi$ restricted to $\mathcal{V}$ are characterized as
projections of the solutions of Lagrange's system for which a bi-homogeneous
structure is exhibited. Secondly we prove a bi-homogeneous B\'ezout Theorem,
which bounds the sum of the degrees of the equidimensional components of the
radical of an ideal generated by a bi-homogeneous polynomial family. This
result is improved when $(f_1, \ldots, f_s)$ is a regular sequence. Moreover,
we use Lagrange's system to design an algorithm computing at least one point
in each connected component of a smooth real algebraic set. This algorithm
generalizes, to the non equidimensional case, the one of Safey El Din and
Schost. The evaluation of the output size of this algorithm gives new upper
bounds on the first Betti number of a smooth real algebraic set. Finally, we
estimate its arithmetic complexity and prove that in the worst cases it is
polynomial in $n$, $s$, $D^s(D-1)^{n-s}{\binom{n}{n-s}}$ and the complexity of
evaluation of $f_1, \ldots, f_s$.}
\RRmotcle{calcul formel, r\'esolution de syst\`emes polynomiaux, g\'eom\'etrie
  alg\'ebrique r\'eelle effective}
\RRkeyword{computer algebra, polynomial system solving, effective real
  algebraic geometry}
\RRprojets{SALSA}
\RRtheme{\THSym} 
\URRocq 

\usepackage{amssymb, amsmath}
\def\Pf{\mathfrak{P}}

\def\Pc{\ensuremath{\mathcal{P}}}
\def\Bc{\ensuremath{\mathcal{B}}}
\def\Vc{\ensuremath{\mathcal{V}}}
\def\Uc{\ensuremath{\mathcal{U}}}
\def\Qc{\ensuremath{\mathcal{Q}}}

\def\C {\ensuremath{\mathbb{C}}}
\def\P {\ensuremath{\mathbb{P}}}
\def\Q {\ensuremath{\mathbb{Q}}}
\def\N {\ensuremath{\mathbb{N}}}
\def\R {\ensuremath{\mathbb{R}}}
\def\Z {\ensuremath{\mathbb{Z}}}

\def\Ass{\mathrm{Ass}}

\def\Adm{{\rm Adm}}
\def\bideg{\mbox{{\rm bideg}}}
\def\<{\langle}
\def\>{\rangle}

\def\X{X_1,\ldots,X_n}

\newtheorem{definition}{Definition}
\newtheorem{corollary}{Corollary}
\newtheorem{proposition}{Proposition}
\newtheorem{lemma}{Lemma}
\newtheorem{theorem}{Theorem}
\newtheorem{remark}{Remark}

\newenvironment{demo}{\textit{Proof. }}{\hfill\penalty-10000\hbox{\kern1mm}\hfill$\square$}

\begin{document}
\makeRR   

\section{Introduction}

Consider polynomials $(f_1, \ldots, f_s)$ in $\Q[X_1, \ldots, X_n]$
(with $s\leqslant n-1$) of degree bounded by $D$. Suppose that this
polynomial family generates a radical ideal and defines a smooth
algebraic variety $\mathcal{V}\subset\C^n$. The core of this paper is
to give an optimal bound on the degree of the critical locus of a
projection $\pi:\C^n\rightarrow \C$ restricted to $\mathcal{V}$ and to
provide an algorithm computing at least one point in each connected
component of the real algebraic set $\mathcal{V}\cap \R^n$ whose worst
case complexity is polynomial in this bound.

\paragraph*{Motivation and description of the problem.} Since computing
critical points solves the problem of algebraic optimization, it has
many applications in chemistry, electronics, financial mathematics
(see~\cite{Floudas} for a non-exhaustive list of problems and
applications). More traditionally, computations of critical points are
used in effective real algebraic geometry to compute at least one
point in each connected component of a real algebraic set. Indeed,
every polynomial mapping restricted to a smooth compact real algebraic
set reaches its extrema on each connected component. Thus, computing
the critical locus of such a mapping provides at least one point in
each connected component of a smooth compact real algebraic set.

In~\cite{GV88,HRS89,HRS93,BPR96,BPR98}, the authors consider the
hypersurface defined by $f_1^2+\cdots+ f_s^2=0$ to study the real
algebraic set $\mathcal{V}\cap\R^n$. The problem is reduced, via
several infinitesimal deformations, to a smooth and compact
situation. Such techniques yield algorithms returning zero-dimensional
algebraic sets encoded by rational parameterizations of degree
$\mathcal{O}(D)^n$ (the best bound is obtained in~\cite{BPR96,BPR98}
and is $(4D)^n$). Similar techniques based on the use of a distance
function to a generic point and a single infinitesimal deformation
(see~\cite{rouillier00a}) yield the bound $(2D)^n$ on the output.

More recently, other algorithms, avoiding the sum of squares (and the
associated growth of degree) have been proposed
(see~\cite{aubry00b,safey_el_din01a,BGHM3,BGHM2,BGHM1,sasc02,sasc03}). The
compactness assumption is dropped either by considering distance
functions and their critical locus
(see~\cite{aubry00b,safey_el_din01a,BGHM3,BGHM4}) or by ensuring
properness properties of some projection functions
(see~\cite{sasc03}). These algorithms compute the critical loci of
polynomial mappings restricted to {\em equidimensional} and smooth
algebraic varieties of dimension $d$, defined by polynomial systems
generating radical ideals. Indeed, under these assumptions, critical
points can be algebraically defined as points where the jacobian
matrix has rank $n-d$, and thus by the vanishing of some minors of the
considered jacobian matrix. On the one hand, some of these algorithms
allow us to obtain efficient implementations (see~\cite{raglib}) while
the algorithms mentioned in the above paragraph do not permit to
obtain usable implementations. On the other hand, applying the
classical B\'ezout bound to the polynomial systems defining the
critical locus of a projection provides a degree bound on the output
which is equal to $D^{n-d}\left ((n-d)(D-1)\right )^d$
(see~\cite{sasc03,BGHM3} where such a bound is explicitly
mentioned). This bound is worse than the aforementioned bounds, but
it has never been reached in the experiments we performed with our
implementations.

Remark that these polynomial systems defining critical points are not
generic: they are overdetermined, and the extracted minors from the
jacobian matrix depend on $f_1, \ldots, f_s$, so that one can hope
that the classical B\'ezout bound is pessimistic.

Our aim is twofold:
\begin{itemize}
  \item providing an optimal bound (in the sense that it can be reached) on
  the degree of the critical locus of a polynomial mapping restricted to an
  algebraic variety;
  \item and designing an algorithm computing at least one point in
  each connected component of a real algebraic set whose worst case
  complexity is polynomial in this bound.
\end{itemize}

\paragraph*{Main contributions.} Consider a projection $\pi$ from $\C^n$ 
onto a line whose base-line vector is $\mathbf{e}\in \C^n$, and the
restriction of $\pi$ to the smooth algebraic variety $\mathcal{V}\subset
\C^n$.  {\em Lagrange's characterization} of critical points of $\pi$
restricted to $\mathcal{V}$ consists in writing that, at a critical point,
there exists a linear relation between the vectors $(\mathbf{grad}(f_1),
\ldots, \mathbf{grad}(f_s), \mathbf{e})$. The resulting polynomial system is
called in the sequel {\em Lagrange's system}. Additional variables are
introduced and are classically called {\em Lagrange multipliers}. Equipped
with such a characterization, critical points can be geometrically seen as
{\em projections of the complex solution set of Lagrange's system}. We prove
that such a characterization remains valid {\em even in non equidimensional
situations}, contrary to the algebraic characterization of
critical points used
in~\cite{aubry00b,safey_el_din01a,BGHM3,BGHM2,BGHM1,sasc02,sasc03}. If $f_1,
\ldots, f_s$ is a regular sequence and if the critical locus of $\pi$
restricted to $\mathcal{V}$ is zero-dimensional, then Lagrange's system is a
zero-dimensional ideal, else it is a positive dimensional ideal.

Since Lagrange multipliers appear with degree $1$ in Lagrange's
system, bounding the degree of the critical locus of $\pi$ restricted
to $\mathcal{V}$ is equivalent to bounding the sum of the degrees of
the isolated primary components of the ideal generated by Lagrange's
system. Lagrange's system can be easily transformed into a
bi-homogeneous polynomial system by a bi-homogenization process which
distinguish the variables $X_1, \ldots, X_n$ from the Lagrange
multipliers. Thus, bounding the degree of a critical locus is reduced
to proving a strong bi-homogeneous B\'ezout Theorem, i.e. to proving a
bound on the {\em sum of the degrees of all the isolated primary
components defining a non-empty bi-projective variety} of an ideal
generated by a bi-homogeneous system. In the sequel, this sum is
called the strong bi-degree of a bi-homogeneous ideal. Such a result
is obtained by using a convenient notion of {\em bi-degree}, which is
given originally in~\cite{VdW78}. In~\cite{VdW78}, a multi-homogeneous
B\'ezout theorem is proved and provides a bound on the sum of the
degrees of the isolated primary components of {\em maximal dimension},
which is not sufficient to reach our goal.

We generalize this result by proving that the same quantity bounds the
sum of the degrees of all the isolated primary components defining a
bi-projective variety of an ideal defined by a bi-homogeneous
polynomial system (see Theorem~\ref{bezout}). Additionally, we prove
that the strong bi-degree of an ideal generated by a bi-homogeneous
polynomial system equals the strong degree of this ideal augmented
with two generic homogeneous affine linear forms lying respectively in
each block of variables (see Theorem~\ref{thm:affine}).

This allows us to prove that the critical locus of a projection $\pi:
\C^n\rightarrow \C$ restricted to $\mathcal{V}$ has degree $D^s
(D-1)^{n-s}{\binom{n}{n-s}}$. This bound becomes $D^s
(D-1)^{n-s}{\binom{n-1}{n-s}}$ when $(f_1, \ldots, f_s)$ is a regular
sequence (see Theorem~\ref{thm:critique}). Some computer simulations
show it is a sharp estimation, since the bound is reached on several
examples.

Next, we use the aforementioned properties of Lagrange's systems to
generalize, to non equidimensional situations, the algorithm due to
Safey El Din and Schost (see~\cite{sasc03}) which computes at least
one point in each connected component of a smooth and equidimensional
real algebraic set (see Theorem~\ref{thm:algo}). Then, the estimation
of the output size of this generalized algorithm provides some
improved upper bounds on the first Betti number of a smooth real
algebraic set (see Theorem~\ref{boundsasc}).

The complexity of our algorithm depends on the complexity of the routine used
to perform algebraic elimination. We consider the elimination subroutine
of~\cite{lecerf2002} which inherits of
\cite{GiHeMoPa95,GiHaHeMoMoPa97,GiHeMoMoPa98,GiLeSa01}. The procedure of
\cite{lecerf2002} comnputes generic fibers of each equidimensional components
of an algebraic variety defined by a polynomial system provided as input. It
is polynomial in the evaluation complexity of the input system, and in an
intrinsic geometric degree. This allows us to prove that the worst case
complexity of our algorithm is polynomial in $n$, $s$, the evaluation
complexity $L$ of $(f_1, \ldots, f_s)$ and the bi-homogeneous B\'ezout bound
$D^s(D-1)^{n-s}{\binom{n}{n-s}}$ (see Theorem~\ref{thm:complexity}).

\paragraph*{Organization of the paper.} The paper is organized as 
follows. In Section~\ref{sec:bihomo}, we prove the strong
bi-homogeneous B\'ezout Theorem. Additionally, we prove that the
strong bi-degree of an ideal equals the strong degree of the same
ideal augmented with two generic affine linear forms lying in each
block of variables. In Section~\ref{sec:crit}, we focus on the
properties of Lagrange's system and use our B\'ezout Theorem to prove
some bounds on the degree of critical loci of projections on a
line. In Section~\ref{sec:sasc}, we generalize the algorithm provided
in~\cite{sasc03} to the non equidimensional case. Moreover, using the
results of the preceding sections, we provide some improved upper
bounds on the first Betti number of a smooth real algebraic set. The
last section is devoted to the complexity estimation of our algorithm.

\paragraph*{Acknowledgments.} We thank D. Lazard, J. Heintz and the anonymous
referees for their helpful remarks which have allowed us to improve and
correct some mistakes in the preliminary version of this paper.

\section{Strong Bi-homogeneous B\'ezout theorem}\label{sec:bihomo}

This section is devoted to the proof of the strong bi-homogeneous
B\'ezout Theorem. This one generalizes the statements
of~\cite{VdW29,VdW78,Rem01a,Rem01b,MHB}. 

In the first paragraph, we provide some useful properties of
bi-homogeneous ideals which allow to define the notions of bi-degree
and strong bi-degree of a bi-homogeneous ideal, these notion seems to
go back to \cite{VdW29}. In the second paragraph, we relate this
bi-degree with some properties of the Hilbert bi-series of a
bi-homogeneous ideal in the case when it defines a zero-dimensional
bi-projective variety. Such properties are already given in
\cite{VdW29}. In the third paragraph, we provide a canonical form of
Hilbert bi-series. Our statements generalize slightly those of
\cite{Rem01a,Rem01b} and allow us to relate the Hilbert bi-series of a
bi-homogeneous ideal and the Hilbert series of this ideal, seen as a
homogeneous one. In the third paragraph, we investigate how the
bi-degree of a bi-homogeneous ideal augmented with a bi-homogeneous
polynomial is related to the bi-degree of the first considered
ideal. Finally, we prove that, under some assumptions, the classical
bi-homogeneous B\'ezout bound is greater than or equal to the sum of
the bi-degrees of the prime ideals associated to the studied
bi-homogeneous ideal. This generalizes the results of
\cite{VdW29,Rem01a,Rem01b,MHB}. Additionally, we prove that this
quantity is greater than or equal to the sum of the degrees of the
primary ideals associated to the studied bi-homogeneous ideal
augmented with two affine linear forms which generalizes the results
of~\cite{VdW78,Rem01b}. For completeness and readibility, we give full
proofs of the required intermediate results.

We denote by $R$ the polynomial ring $\Q[X_0, \ldots, X_n, \ell_0,
\ldots,\ell_k]$.


\subsection{Preliminaries and main results}\label{sec:prelim}

In this paragraph, we introduce some basics on bi-homogeneous
ideals. We prove that given a bi-homogeneous ideal $I\subset R$, there
exists a minimal primary bi-homogeneous decomposition of $I$, i.e. a
primary decomposition such that each primary component is a
bi-homogeneous ideal. We define a notion of {\em bi-degree} and {\em
strong bi-degree} of bi-homogeneous ideals. Finally, we state the main
results (see Theorems $1$ and $2$ below) of the section which bound
the strong bi-degree of a bi-homogeneous ideal $I$ by the classical
bi-homogeneous B\'ezout bound on the one hand, and show that the {\em
strong bi-degree} of $I$ bounds the degree of $I+\<u-1, v-1\>$ (where
$u$ and $v$ are homogeneous linear forms respectively chosen in
$\Q[X_0, \ldots, X_n]$ and $\Q[\ell_0, \ldots, \ell_k]$).

\begin{definition}\label{def:bihompolideal}
A linear form in $R$ is a polynomial of degree $1$ whose support contains only
monomials of degree at most $1$. A homogeneous linear form is a linear form
whose support contains {\em only} monomials of degree $1$.

A polynomial $f$ in $R$ is said to be bi-homogeneous if and only if there
exists a {\em unique} couple of integers $(\alpha,\beta)$ such that for all
$(u, v)\in \Q\times\Q$:
$$
f(uX_0, \ldots, uX_n,v\ell_0, \ldots, v\ell_k)=u^\alpha v^\beta f(X_0, \ldots, X_n,
\ell_0, \ldots, \ell_n).
$$
The couple $(\alpha, \beta)$ is called the {\em bi-degree} of
$f$.

Given a polynomial $f\in R$, the bi-homogeneous component of bi-degree
$(\alpha,\beta)$ of $f$, denoted by $f_{\alpha,\beta}$, is the unique
bi-homogeneous polynomial such that the support of
$f-f_{\alpha,\beta}$ does not contain any monomial of bi-degree
$(\alpha,\beta)$.

An ideal $I\subset R$ is said to be a {\em bi-homogeneous} ideal if and only
if for all $f\in I$, and for all bi-degree $(\alpha,\beta)$,
$f_{\alpha,\beta}\in I$.
\end{definition}

\begin{lemma}\label{lemme:bihomgen}
Let $I\subset R$ be a {\em bi-homogeneous} ideal. Then, there exists a finite
polynomial family $f_1, \ldots, f_s$ generating $I$ such that each $f_i$
(for $i=1, \ldots, s$) is a bi-homogeneous polynomial. 
\end{lemma}

\begin{demo}
Consider a finite set of generators $\mathcal{F}$ of $I$ (there exists
one since $R$ is Noetherian). Since $I$ is a bi-homogeneous ideal, the
{\em finite} set $\widetilde{\mathcal{F}}$ of the bi-homogeneous
components of all the polynomials in $\mathcal{F}$ generates an ideal
$J$ which is contained in $I$.

Consider now $f\in I$. Since $I$ is generated by $\mathcal{F}$ and since each
polynomial of $\mathcal{F}$ is a linear combination of some polynomials in
$\widetilde{\mathcal{F}}$, $I$ is contained in $J$.
\end{demo}


\begin{lemma}\label{lemme:primhom}
Let $Q_1, \ldots, Q_p$ be a family of bi-homogeneous ideals in $R$. Then
$Q_1\cap\cdots\cap Q_p$ is a bi-homogeneous ideal.   
\end{lemma}

\begin{demo}
Denote by $I$ the ideal $Q_1\cap\cdots\cap Q_p$ and consider $f\in I$. For all
$i\in\{1, \ldots, p\}$, $f\in Q_i$ and $Q_i$ is, by assumption,
bi-homogeneous. Then, for all $i\in\{1, \ldots, p\}$, each bi-homogeneous
component of $f$ belongs to $Q_i$, which implies that each bi-homogeneous
component belongs to $I$.   
\end{demo}

Given a bi-homogeneous ideal $I\subset R$, we now prove that there exists a
primary decomposition of $I$ for which each primary component is
bi-homogeneous.

\begin{proposition}\label{prop:decbihomid}
  Let $I$ be a bi-homogeneous ideal, and $Q_1\cap\cdots\cap Q_p$ be a minimal
  primary decomposition of $I$. Then there exist {\em primary} bi-homogeneous
  ideals $Q_1^\prime, \ldots, Q^\prime_p$ such that
  $I=Q_1^\prime\cap\cdots\cap Q^\prime_p$.
\end{proposition}

\begin{demo}
  For $i=1, \ldots, p$, consider $Q_i$, a primary ideal of the above
  minimal primary decomposition of $I$ and let $Q_i^\prime$ be the
  ideal generated by the bi-homogeneous polynomials of $Q_i$. First,
  remark that $Q_i^\prime$ is non empty since it contains $I$. We
  prove now that $Q^\prime_i$ (for $i=1, \ldots, p$) is a primary
  ideal and then that $I=Q^\prime_1\cap\cdots\cap Q^\prime_p$.

  
  Let $f$ and $g$ be two polynomials such that $fg\in Q_i^\prime$ and
  $g\notin Q_i^\prime$. We show below that this implies there exists
  an integer $N$ such that $f^N\in Q_i^\prime$ which, in turn, implies
  $Q_i^\prime$ is a primary ideal. The proof is done by induction on
  the number $h$ of bi-homogeneous components of $f$.

  If $h=1$, $f$ is bi-homogeneous, and the result is obvious. Suppose now that
  for any polynomial $\widetilde{f}$ having $h$ bi-homogeneous components,
  $\widetilde{f}g\in Q_i^\prime$ with $g\notin Q_i^\prime$ implies there
  exists an integer $N$ such that $\widetilde{f}^N\in Q_i^\prime$.


  Consider $f$ having $h+1$ bi-homogeneous components and such that
  there exists $g\notin Q_i^\prime$ with $fg\in Q_i'$. Since
  $Q_i^\prime$ is bi-homogeneous, each bi-homogeneous component of
  $fg$ belongs to $Q_i^\prime$. Remark that there exists one
  bi-homogeneous component of the product $fg$ of maximal degree which
  can be written as a product $f_{\alpha, \beta}.g_{\alpha^\prime,
  \beta^\prime}$ where $f_{\alpha, \beta}$ (resp. $g_{\alpha^\prime,
  \beta^\prime}$) is a bi-homogeneous component of bi-degree $(\alpha,
  \beta)$ (resp. $(\alpha^\prime, \beta^\prime)$) of $f$
  (resp. $g$). Moreover, without loss of generality one can suppose
  $g_{\alpha^\prime, \beta^\prime}\notin Q_i$: if it is not the case,
  it is sufficient to substitute $g$ by $g-g_{\alpha^\prime,
  \beta^\prime}$.

  Since $Q_i$ is a primary ideal, there exists $M\in\N$, such that
  $f_{\alpha,\beta}^{M}\in Q_i$. Moreover, since $f_{\alpha,\beta}$ is
  bi-homogeneous, $f_{\alpha,\beta}^{M}$ is bi-homogeneous. Thus, since
  $Q_i^\prime$ is generated by the bi-homogeneous polynomials of $Q_i$, this
  implies $f_{\alpha,\beta}^{M}\in Q_i'$.

 Since $fg\in Q_i^\prime$, $f^M g\in Q_i^\prime$. Moreover
 $f^Mg=(f-f_{\alpha, \beta})^M g+f_{\alpha,\beta}^M g$ while
 $f_{\alpha,\beta}^{M}\in Q_i'$. This implies
 $(f-f_{\alpha,\beta})^{M}g\in Q_i'$. Suppose now that $M$ is the
 smallest integer such that $(f-f_{\alpha,\beta})^{M}g\in Q_i'$ and
 remark that this implies $(f-f_{\alpha,\beta})^{M-1}g\notin
 Q_i'$. Thus, the above reasoning can be done using $(f-f_{\alpha,
 \beta})$ (which has $h$ bi-homogeneous component) instead of $f$ and
 $\widetilde{g}=(f-f_{\alpha,\beta})^{M-1}g\notin Q_i'$ instead of
 $g$. {F}rom the induction hypothesis, this implies there exists an
 integer $M^\prime$ such that $(f-f_{\alpha,\beta})^{M^\prime}\in
 Q_i^\prime$, while the integer $M$ is such that
 $f_{\alpha,\beta}^{M}\in Q_i^\prime$. Thus, considering the binomial
 development of $(f-f_{\alpha, \beta}+f_{\alpha,\beta})^N$ where
 $N>2M$ shows that $f^N\in Q_i^\prime$.

  
  It remains to prove that $I=Q_1^\prime\cap \cdots\cap Q_m^\prime$. To this
  end, remark that for all $i$, $I\subset Q_i$ and as $I$ is
  bi-homogeneous, $I\subset Q_i'$ while $Q_i^\prime\subset Q_i$. Thus one has
  $I\subset Q_1^\prime \cap\cdots\cap Q_m^\prime\subset Q_1\cap \cdots\cap Q_m
  =I$ which ends the proof.  
\end{demo}

{F}rom now on, given a bi-homogeneous ideal $I$, we {\em only}
consider {\em bi-homoge\-neous} primary decompositions of $I$, i.e.
decompositions of $I$ such that each primary component is
bi-homo\-geneous. Remark that from such a bi-homogeneous primary
decomposition, one can extract a {\em minimal} primary decomposition.
{F}rom the uniqueness of the isolated primary components of a minimal
primary decomposition (see~\cite{Eisenbud}), one deduces the
uniqueness of the isolated primary components of a bi-homogeneous
minimal primary decomposition of $I$. This leads to the following
result.

\begin{corollary}\label{corol:primedecbihom}
Let $I$ be a bi-homogeneous ideal of $R$. There exists a minimal primary
decomposition of $I$ such that each primary component is a bi-homogeneous
ideal. The set of isolated bi-homogeneous components is unique.
\end{corollary}

Note that primary ideals of $R$ which contain a power of $\<X_0, \ldots,
X_n\>$ or a power of $\<\ell_0, \ldots, \ell_k\>$ define an empty
bi-projective variety in $\P^n(\C)\times\P^k(\C)$. 

\begin{definition}\label{def:bihom}
A primary bi-homogeneous ideal is said to be {\em admissible} if and only if
it contains neither a power of $\<X_0, \ldots, X_n\>$ nor a power of
$\<\ell_0, \ldots, \ell_k\>$.

A primary component of a bi-homogeneous ideal is said to be {\em admissible}
if it is an admissible ideal.

The set of admissible isolated bi-homogeneous components of a minimal
bi-homogeneous primary decomposition of an ideal $I\subset R$ is denoted by
${\rm Adm}(I)$.


Let $I\subset R$ be a bi-homogeneous ideal, and $(d, e)$ be a couple in
$\N\times \N$. The couple $(d, e)$ is an {\em admissible bi-dimension} of $I$
if and only if $d\leqslant n$, $e \leqslant k$ and $d+e+2$ equals the maximum of the
Krull dimensions of the ideals in ${\rm Adm}(I)$. 
\end{definition}

\begin{remark}
Consider a bi-homogeneous ideal $I\subset R$ and let $J=\cap_{\mathcal{Q}\in
{\rm Adm}(I)}\mathcal{Q}$. {F}rom Lemma~\ref{lemme:primhom}, $J$ is a
bi-homogeneous ideal.
\end{remark}

A bi-homogeneous ideal $I$ in $R$ defines a non-empty bi-projective variety
$\mathcal{V}$ in $\P^{n}(\C)\times\P^{k}(\C)$ if and only if ${\rm Adm}(I)$ is
not empty. Note also that the maximal dimension of the admissible primary
components of $I$ can be less than the Krull dimension of
$I$.

In the homogeneous context, the degree of an ideal of Krull dimension $D$
defining a non-empty projective variety is obtained as the degree of this
ideal augmented with $D$ {\em generic} linear forms (see \cite{CLO}).  Below, we
provide a similar process to define a notion of bi-degree of a bi-homogeneous
ideal.

In the sequel, a homogeneous linear form $u=\sum_{i=0}^n u_iX_i\in\Q[X_0,
\ldots, X_n]$ (resp. $v=\sum_{j=0}^k v_j\ell_j \in \Q[\ell_0, \ldots,
\ell_k]$) is identified to the point $(u_0, \ldots, u_n)\in \Q^{n+1}$
(resp. $(v_0, \ldots, v_k)\in \Q^{k+1}$).

\begin{proposition}\label{lemme:genericintersection}
Let $I\subset R$ be an ideal of dimension $D\geqslant 2$ and $(d, e)\in\N\times\N$
such that $d+e+2=D$. 
\begin{itemize}
  \item Either for any choice of homogeneous linear forms $(u_1,
\ldots, u_{d+1})$ in $\Q[X_0,\penalty-10000 \ldots, X_n]$ and $(v_1, \ldots, v_{e+1})$ in
$\Q[\ell_0, \ldots, \ell_k]$ the ideal
$$
I+\<(u_1-1), u_2, \ldots, u_{d+1}, (v_1-1), v_2, \ldots, v_{e+1}\>
$$ equals $\Q[X_0, \ldots, X_n, \ell_0, \ldots, \ell_k]$; 
\item or there exist an integer $\mathfrak{D}\in \N$ and a Zariski
  closed subset $\mathcal{H}\subsetneq(\C^{n+1})^{d+1}\times
  (\C^{k+1})^{e+1}$ such that for any choice of homogeneous linear
  forms $(u_1, \ldots, u_{d+1}, v_1, \ldots,\penalty-10000 v_{e+1})$
  outside $\mathcal{H}$ (where for $i\in\{1, \ldots, d+1\}$, $u_i\in \Q[X_0,
  \ldots, X_n]$ and for $j\in\{1, \ldots, e+1\}$, $v_j\in \Q[\ell_0, \ldots,
  \ell_k]$) the ideal
$$
I+\<(u_1-1), u_2, \ldots, u_{d+1}, (v_1-1), v_2, \ldots, v_{e+1}\>
$$
is zero-dimensional and its degree is $\mathfrak{D}$. 
\end{itemize}
\end{proposition}

\begin{demo}
For $i\in\{1, \ldots, d+1\}$, $p\in\{0, \ldots, n\}$, $j\in\{1, \ldots, e+1\}$, and $q\in\{0,
\ldots, k\}$, let $\mathfrak{u}_{i,p}$, and $\mathfrak{v}_{j,q}$ be new
indeterminates. Denote by $\mathfrak{U}$ (resp. $\mathfrak{V}$) the set of
indeterminates $\{\mathfrak{u}_{1, 0}, \ldots, \mathfrak{u}_{d+1, n}\}$
(resp. $\{\mathfrak{v}_{1, 0}, \ldots, \mathfrak{v}_{e+1, k}\}$).

Let $K$ be the field of rational fractions $\Q(\mathfrak{U}, \mathfrak{V})$;
for $i\in \{1, \ldots, d+1\}$, let $\bar{K}_i$ be the field of rational
fractions $\Q(\mathfrak{U}\setminus\{\mathfrak{u}_{i,0}, \ldots,
\mathfrak{u}_{i, n}\}, \mathfrak{V})$ and for $j\in \{1, \ldots, e+1\}$, let
$\underline{K}_j$ be the field of rational fractions $\Q(\mathfrak{U},
\mathfrak{V}\setminus\{\mathfrak{v}_{j,0}, \ldots, \mathfrak{v}_{j,k}\})$.

Consider for $2\leqslant i\leqslant d+1$ (resp. $2\leqslant j\leqslant e+1$) the linear forms
$\mathfrak{u}_i=\sum_{p=0}^n\mathfrak{u}_{i,p}X_p$
(resp. $\mathfrak{v}_j=\sum_{q=0}^k\mathfrak{v}_{j,q}\ell_q$) and
$\mathfrak{u}_{1}=\sum_{p=0}^k\mathfrak{u}_{1,p}X_p-1$
(resp. $\mathfrak{v}_{1}=\sum_{q=0}^k\mathfrak{v}_{1,q}\ell_q-1$).

For $i\in\{1, \ldots, d+1\}$ (resp. $j\in\{1, \ldots, e+1\}$)and $u=(u_0, \ldots, u_n)$
(resp. $v=(v_0, \ldots, v_k)$) a point in $\Q^{n+1}$ (resp. $\Q^{k+1}$),
denote by $\varphi_{i,u}$ (resp. $\psi_{j,v}$) the ring homomorphism
$\varphi_{i,u} : K[X_0, \ldots, X_n, \ell_0, \ldots, \ell_k]\rightarrow
\bar{K}_i[X_0, \ldots, X_n, \ell_0, \ldots, \ell_k]$ (resp. $\psi_{j,v} :
K[X_0, \ldots,X_n, \ell_0, \ldots, \ell_k]\rightarrow \underline{K}_j[X_0,
\ldots, X_n, \ell_0, \ldots, \ell_k]$) such that:
\begin{itemize}
  \item for all $p\in\{0, \ldots, n\}$ (resp. $q\in\{0, \ldots,
    k\}$),$\varphi_{i,u}(\mathfrak{u}_{i,p})=u_{p}$
    (resp. $\psi_{j,v}(\mathfrak{v}_{j,q})=v_{q}$),
  \item for all $p\in\{0, \ldots, n\}$ (resp. $q\in\{0, \ldots, k\}$), and $r\in \{1,
    \ldots, d+1\}\setminus\{i\}$ (resp. $s\in \{1, \ldots,
    e+1\}\setminus\{j\}$), $\varphi_{i,u}(\mathfrak{u}_{r,p})=\mathfrak{u}_{r,p}$
    (resp. $\psi_{j,v}(\mathfrak{v}_{s,q})=\mathfrak{v}_{s,q}$),
  \item $\varphi_{i,u}(X_p)=X_p$ (resp. $\psi_{j,v}(X_p)=X_p$) and
$\varphi_{i,u}(\ell_q)=\ell_q$ (resp. $\psi_{j,v}(\ell_q)=\ell_q$).
\end{itemize}

Finally, given a couple of points $u=(u_{1,0}, \ldots, u_{1,n}, \ldots,
u_{d+1, 0}, \ldots, u_{d+1, n})$ (resp. $v=(v_{1,0}, \ldots,
v_{1,k}, \ldots, v_{e+1, 0}, \ldots, v_{e+1, k})$) in $(\Q^{n+1})^{d+1}$
(resp. $(\Q^{k+1})^{e+1}$), denote by $\vartheta_{(u,v)}$ the ring
homomorphism $\vartheta_{(u,v)} : K[X_0, \ldots, X_n, \ell_0,
\ldots, \ell_k]\rightarrow \Q[X_0, \ldots, X_n, \ell_0, \ldots, \ell_k]$ such
that for all $p=0, \ldots, n$ and $q=0, \ldots, k$),
$\vartheta_{u,v}(\mathfrak{u}_{i,p})= u_{i,p}$,
$\vartheta_{u,v}(\mathfrak{v}_{j,q})=v_{j,q}$,
$\vartheta_{u,v}(X_p)=X_p$ and $\vartheta_{u,v}(\ell_q)=\ell_q$.

In the sequel, $J_0=I$, for $i\in \{1, \ldots, d+1\}$, $J_i$ is the ideal
$I+\<\mathfrak{u}_{1}, \ldots, \mathfrak{u}_{i}\>$, and for $i\in
\{d+2,\ldots, D\}$, $J_i$ is the ideal $I+\<\mathfrak{u}_{1}, \ldots,
\mathfrak{u}_{d+1}, \mathfrak{v}_{1}, \ldots, \mathfrak{v}_{i-(d+1)}\>$.

We first prove that the ideal $J_D$ is either zero-dimensional in $K[X_0,
\ldots, X_n, \penalty-10000\ell_0, \ldots, \ell_k]$ or equals $K[X_0, \ldots,
X_n, \ell_0, \ldots, \ell_k]$. The proof is done by proving that, for $i=1,
\ldots, D$, $\dim(J_i) < \dim(J_{i-1})$.

Consider the case $i=1$ and suppose that $\dim(J_1)\geqslant \dim(J_0)>0$. This
implies that there exists an isolated primary component $\mathcal{Q}_0$ of
$J_0$ {\em of dimension $\dim(J_0)>0$} and an integer $N$ such that
$\mathfrak{u}_{1}^N\in \mathcal{Q}_0$. 

Since $J_0$ is generated by a polynomial family which does not involve
the indeterminates $\mathfrak{u}_{1, 0}, \ldots, \mathfrak{u}_{1, n}$,
for any isolated primary component $\mathcal{Q}$ of $J_0$,
$\mathcal{Q}$ is generated by a polynomial family which does not
involve the indeterminates $\mathfrak{u}_{1, 0}, \ldots,
\mathfrak{u}_{1, n}$. Thus, for any $u\in \Q^{n+1}$,
$\varphi_{1,u}(\mathcal{Q}_0)=\mathcal{Q}_0\cap \bar{K}_1$.  Then, there exists a
Zariski-closed subset $\mathcal{Z}\subset \C^{n+1}$ such that for any
point $u$ in $\Q^{n+1}\setminus \mathcal{Z}$,
$\varphi_{1,u}(\mathfrak{u}_1)^N$ belongs to $\mathcal{Q}_0$. Choose
now $n+2$ points $u_1, \ldots, u_{n+2}$ in $\Q^{n+2}$ such that
$\<\varphi_{1,u_1}(\mathfrak{u}_{1}^N), \ldots, \varphi_{1,
u_{n+2}}(\mathfrak{u}_{1}^N)\>=\<1\>$.

Since $\<\varphi_{1,u_1}(\mathfrak{u}_{1}^N), \ldots, \varphi_{1,
u_{n+2}}(\mathfrak{u}_{1}^N)\>\subset\mathcal{Q}_0$, $\mathcal{Q}_0=K[X_0,
\ldots, X_n, \ell_0, \ldots, \ell_k]$ which contradicts
$\dim(\mathcal{Q}_0)=D>0$. Thus, $\dim(J_1)<\dim(J_0)$.

Consider now the case where $2\leqslant i \leqslant d+1$ and suppose that $\dim(J_i)\geqslant
\dim(J_{i-1})>0$. This implies that there exists an isolated primary component
$\mathcal{Q}_0$ of $J_{i-1}$ {\em of positive dimension} and an integer $N$ such
that $\mathfrak{u}_{i}^N\in \mathcal{Q}_0$.

Since $J_{i-1}$ is generated by a polynomial family which does not involve the
indeterminates $\mathfrak{u}_{i, 0}, \ldots, \mathfrak{u}_{i, n}$,
$\mathcal{Q}_0$ is generated by a polynomial family which does not involve the
indeterminates $\mathfrak{u}_{i, 0}, \ldots, \mathfrak{u}_{i, n}$ and thus for
any $u\in \Q^{n+1}$, $\varphi_{i,u}(\mathcal{Q}_0)=\mathcal{Q}_0\cap\bar{K}_i$.  Then, for
any point $u$ in $\Q^{n+1}$, $\varphi_{i,u}(\mathfrak{u}_i)^N$ belongs to
$\mathcal{Q}_0$. Choosing $n+2$ points $u_1, \ldots, u_{n+2}$ in $\Q^{n+2}$
such that $\<\varphi_{i,u_1}(\mathfrak{u}_{i}^N),
\ldots,\varphi_{i,u_{n+2}}(\mathfrak{u}_{i}^N)\>=\<X_0^N, \ldots,
X_n^N\>\subset\mathcal{Q}$. Since $\mathfrak{u}_{1}\in
J_{i-1}\subset\mathcal{Q}_0$ and $\<\mathfrak{u}_{1}\>+\<X_0^N, \ldots,
X_n^N\>=\<1\>$ this implies that $\mathcal{Q}_0=K[X_0, \ldots, X_n, \ell_0,
\ldots, \ell_k]$ which contradicts the fact that $\mathcal{Q}_0$ has positive
dimension. Thus, $\dim(J_i)<\dim(J_{i-1})$.

Proving that $\dim(J_{d+2})<\dim(J_{d+1})$ is done by the same way as proving
that $\dim(J_1)<\dim(J_0)$ using the homomorphism $\psi_{d+2, v}$ instead of
$\varphi_{1, u}$ (for $u\in \Q^{n+1}$ and $v\in \Q^{k+1}$). Proving for
$i>d+2$ that $\dim(J_i)< \dim(J_{i-1})$ is done following the same arguments
as those of the above paragraph (using the homomorphisms $\psi_{i, v}$ for
$v\in \Q^{k+1}$).

Thus, if $J_D\neq K[X_0, \ldots, X_n , \ell_0, \ldots, \ell_k]$, one has
$\dim(J_D) < \dim(J_{D-1}) <\cdots<\dim(J_i)<\dim(J_{i-1}) < \cdots < \dim(I)$
which implies $J$ is be zero-dimensional.

If $J_D=K[X_0, \ldots, X_n, \ell_0, \ldots, \ell_k]$, for any $(u,v )\in
(\Q^{n+1})^{d+1}\times (\Q^{k+1})^{e+1}$, $\vartheta_{u,v}(J_D)$ equals
$\Q[X_0, \ldots, X_n]$.

Suppose now $J_D$ to be zero-dimensional and consider a Gr\"obner basis $G$ of
$J_D$. Let $\mathcal{H}\subset (\C^{n+1})^{d+1}\times(\C^{k+1})^{e+1}$ be the
Zariki-closed set which is the union of zero-sets of the common denominator of
each polynomial of $G$. Thus, for any point $(u, v)\in (\Q^{n+1})^{d+1}\times
(\Q^{k+1})^{e+1}\setminus\mathcal{H} $, $\vartheta_{u,v}(G)$ is a Gr\"obner
basis of $\vartheta_{u,v}(J_D)$ (see~\cite{Gianni}). Then,
$\vartheta_{u,v}(J_D)$ is zero-dimensional and its degree is the one of
$J_D$. 
\end{demo}

\begin{remark}\label{remark:dim2}
{F}rom the proof of the above Proposition, one deduces also the following:
\begin{itemize}
  \item Let $I\subset R$ be a bi-homogeneous ideal which is an intersection of
  admissible primary ideals. Then, there exists a Zariski-closed subset
  $\mathcal{H}\subsetneq\C^{n+1}\times \C^{k+1}$ such that for all $(u_1,
  v_1)\in\C^{n+1}\times \C^{k+1}\setminus \mathcal{H}$, $u_1-1$ and $v_1-1$ do
  not divide $0$ in $R/I$ and $v_1-1$ does not divide $0$ in $R/I+\<u_1-1\>$. 

  Moreover if $I$ is equidimensional, $I+\<u_1-1, v_1-1\>$ is
  equidimensional. 
  \item Let $I\subset R$ be an equidimensional bi-homogeneous ideal which is
  an intersection of admissible primary ideals and $J=I+\<u_1-1, v_1-1\>$. If
  $J\neq R$, there exists $(d,e)\in \N\times \N$ such that $d+e=\dim(J)$ and a
  Zariski-closed subset
  $\mathcal{H}\subsetneq(\C^{n+1})^{d}\times(\C^{k+1})^{e+1}$ such that for
  any choice of homogeneous linear forms $(u_2, \ldots, u_{d+1}, v_2, \ldots,
  v_{e+1})\in \Q[X_0, \ldots, X_n]\times\Q[\ell_0, \ldots, \ell_k]\setminus
  \mathcal{H}$, $u_2$ does not divide $0$ in $R/J$, $u_i$ does not divide $0$
  in $R/J+\<u_2, \ldots, u_{i-1}\>$ (for $i=3, \ldots, d+1$), $v_2$ does not
  divide zero in $R/J+\<u_2, \ldots, u_{d+1}\>$ and $v_j$ does not divide $0$
  in $R/J+\<u_2, \ldots, u_{d+1}, v_1, \ldots, v_{j-1}\>$ (for $j\in \{3,
  \ldots, e+1\}$).
\end{itemize}
\end{remark}

In the sequel, we shall say that a property $\mathfrak{P}$ is true for a {\em
generic} choice of linear forms, if there exists a Zariski-closed subset in
the set of the considered linear forms such that for any choice of forms
outside this Zariski-closed subset, the property $\mathfrak{P}$ is satisfied.

Proposition~\ref{lemme:genericintersection} and the uniqueness of ${\rm
Adm}(I)$ allows us to define the following notion of {\em bi-degree} of a
bi-homogeneous ideal $I\subset R$.

\begin{definition}\label{def:bidegre}
Let $I\subset R$ be a bi-homogeneous ideal of dimension $D$. For $(d,
e)\in \N\times \N$ such that $d+e+2=D$, consider the linear forms
$u_1, \ldots, u_{d+1}$ (resp. $v_1, \ldots, v_{e+1}$) which are chosen
{\em generically} in $\Q[X_0, \ldots, X_n]$
(resp. $\Q[\ell_0,\penalty-10000 \ldots, \ell_k]$).  
$C_{d,e}(I)$ denotes the degree of $I+\<u_1,\ldots, u_{d+1}-1, v_1, \ldots,
v_{e+1}-1\>$.

If $I$ is primary, the {\em bi-degree} of $I$ is the sum $\sum_{d+e+2=D} C_{d,e}(I)$. 

If $I$ is not primary, the {\em bi-degree} of $I$ is the sum of the bi-degrees
of the ideals of ${\rm Adm}(I)$ having {\em maximal Krull dimension}. 

If $I$ is not primary, the {\em strong bi-degree} of $I$ is the sum of the
bi-degrees of the ideals of ${\rm Adm}(I)$ (which are isolated by
definition of $\Adm(I)$.
\end{definition}

\begin{remark}\label{rem:bidegadm}
  Let $I\subset R$ be a bi-homogeneous ideal and $J=\cap_{\mathcal{Q}\in {\rm
  Adm}(I)}\mathcal{Q}$ which is bi-homogeneous from
  Lemma~\ref{lemme:primhom}. Note that, by definition, the bi-degree
  (resp. the strong bi-degree) of $I$ equals the bi-degree (resp. strong
  bi-degree) of $J$.
\end{remark}

We are now ready to state the main results of this section. The first one
bounds the strong bi-degree of a bi-homogeneous ideal $I\subset R$ under some
assumptions. This generalizes the statements of~\cite{VdW29,Shafarevich77,MHB}
which only consider with the admissible primary components of ${\rm Adm}(I)$
having maximal dimension.

\begin{theorem}\label{bezout}
 Let $s\in \{1, \ldots, n+k\}$ and $f_1,\ldots,f_s$ be
  bi-homogeneous polynomials in $R$ of respective bi-degree
  $(\alpha_i,\beta_i)$ generating a bi-homogeneous ideal $I$.  Suppose
  that there exist at most $n$ $f_i$ such that $\beta_i=0$ and at most
  $k$ $f_i$ such that $\alpha_i=0$, then the sum of the bi-degrees of
  the bi-homogeneous associated primes of $I$ is bounded by
  $\displaystyle\mathcal{B}(f_1,\ldots,f_s)=\sum_{\mathcal{I},
  \mathcal{J}} \left (\Pi_{i\in\mathcal{I}}\alpha_{i}\right ).\left
  (\Pi_{j\in\mathcal{J}}\beta_{j}\right ) $ where $\mathcal{I}$ and
  $\mathcal{J}$ are disjoint subsets for which the union is $\{1,
  \ldots, s\}$ such that the cardinality of $\mathcal{I}$
  (resp. $\mathcal{J}$) is bounded by $n$ (resp. $k$).
  
\end{theorem}

The following result generalizes the one of~\cite{VdW78} and states
that given an ideal $I\subset\Q[X_1, \ldots, X_n, \ell_1, \ldots,
\ell_k]$ its strong degree (in the meaning of~\cite{Heintz83}) is
bounded by the strong bi-degree of a bi-homogeneous ideal constructed
from a bi-homogeneization process applied to each polynomial in $I$.

\begin{theorem}\label{thm:affine}
Consider the mapping :
$$
\phi\; :\; \begin{array}[t]{ccl}
    \Q[X_1, \ldots, X_n, \ell_1, \ldots, \ell_k]&\rightarrow &  \Q[X_0,X_1, \ldots, X_n, \ell_0,\ell_1, \ldots, \ell_k]\\
      f&\mapsto& X_0^{\deg_X(f)}\ell_0^{\deg_\ell(f)} f(\frac{X_1}{X_0}, \ldots,
      \frac{X_n}{X_0}, \frac{\ell_1}{\ell_0}, \ldots,
      \frac{\ell_k}{\ell_0}) \\
  \end{array}
$$ where $\deg_X(f)$ (resp. $\deg_\ell(f)$) denotes the degree of $f$
seen as a polynomial in $\Q(\ell_1, \ldots, \ell_k)[X_1, \ldots, X_n]$
(resp. $\Q(X_1, \ldots, X_n)[\ell_1, \ldots, \ell_k]$). 

Given an ideal $I\subset\Q[X_1, \ldots, X_n, \ell_1, \ldots, \ell_k]$,
denote by $\phi(I)$ the ideal $\{\phi(f)\mid f\in I\}\subset \Q[X_0,
\ldots, X_n, \ell_0, \ldots, \ell_k]$. 

Then, $\phi(I)$ is a bi-homogeneous ideal and the sum of the degrees
of the isolated primary components of $I$ is bounded by the strong
bi-degree of $\phi(I)$. 
\end{theorem}

The proof of these results rely on the study of Hilbert bi-series of
bi-homogene\-ous ideals which are introduced in~\cite{VdW29}. 

\subsection{Hilbert bi-series: basic properties}\label{sec:biserhilb}

We follow here~\cite{VdW29} which introduce Hilbert bi-series of
bi-homogeneous ideals and study their properties when the considered
bi-homogeneous ideal has bi-dimension $(0,0)$.
\paragraph*{Notation.}
Given a couple $(i, j)\in \N\times\N$, we denote by $R_{i,j}$ the $\Q$-vector
space of the {\em bi-homogeneous} polynomials in $R$ of degree $i$ in the set
of variables $X_0, \ldots, X_n$ and $j$ in the set of variables $\ell_0,
\ldots, \ell_k$.

Given a {\em bi-homogeneous} ideal $I\subset R$ and a couple
$(i,j)\in\N\times\N$, we denote by $I_{i,j}$ the intersection of $I$ with
$R_{i,j}$.

Given a couple $(i, j)\in \N\times\N$, we denote by $R_{\leqslant i,\leqslant j}$ the
$\Q$-vector space of polynomials in $R$ of degree less than or equal to $i$
(resp. $j$) in the set of variables $X_0, \ldots, X_n$ (resp. $\ell_0, \ldots,
\ell_k$).

Given an ideal $I\subset R$ and a couple $(i,j)\in\N\times\N$, we denote by
$I_{\leqslant i,\leqslant j}$ the intersection of $I$ with $R_{\leqslant i,\leqslant j}$.

\begin{definition}\label{def:biseries}
The Hilbert bi-series of a bi-homogeneous ideal $I\subset R$ is the series
$\sum_{i,j}\dim\left (R_{i,j}/I_{i,j}\right ) t_1^i t_2^j$.

The affine Hilbert bi-series of an ideal $I\subset R$ is $\sum_{i,j}\dim\left
(R_{\leqslant i,\leqslant j}/I_{\leqslant i,\leqslant j}\right ) t_1^i t_2^j$.
\end{definition}

Consider a bi-homogeneous ideal $I\subset R$ of Krull dimension $2$. This
section is devoted to prove that, there exists $(i_0,j_0)\in \N\times\N$ such
that for all {\em $i\geqslant i_0$ and $j\geqslant j_0$},
$\dim(R_{i,j}/I_{i,j})=\dim(R_{i_0,j_0}/I_{i_0,j_0})$. Such a result already
appears in \cite{VdW29}.

Denote by $R^\prime$ the polynomial ring $\Q[X_1, \ldots, X_{n}, \ell_1,
\ldots, \ell_{k}]$. Consider the application $\phi_{i,j}: R^\prime_{\leqslant i,
\leqslant j}\rightarrow R_{i,j}$ sending a polynomial $f\in R^\prime_{\leqslant i, \leqslant
j}$ of degree $\alpha$ (resp. $\beta$) in the variables $X_1, \ldots, X_n$
(resp. $\ell_1, \ldots, \ell_k$) to the polynomial
$\phi_{i,j}(f)=X_0^{i}\ell_0^{j} f(\frac{X_1}{X_0}, \ldots,
\frac{X_{n}}{X_0}, \frac{\ell_1}{\ell_0}, \ldots,
\frac{\ell_{k}}{\ell_0})$.

Gi\-ven an ideal $I\subset R^\prime$ and $(i, j)\in \N\times \N$, $\phi_{i,
j}(I)$ denotes the set $\{\phi_{i, j}(f)\mid f\in I_{\leqslant i, \leqslant j}\}$.

Additionally, consider the mapping $\psi_{i, j}: R_{i, j}\rightarrow
R^\prime_{\leqslant i, \leqslant j}$ (which takes place of a dehomogenization process
in a homogeneous context) sending a bi-homoge\-neous polynomial $f\in R_{i, j}$
to $$\psi_{i, j}(f)=f(1, X_1, \ldots, X_{n}, 1, \ell_1, \ldots, \ell_{k})\in
R^\prime.$$

Gi\-ven a bi-homogeneous ideal $I\subset R$ and $(i, j)\in \N\times \N$,
$\psi_{i, j}(I)$ denotes the set $\{\psi_{i, j}(f)\mid f\in I_{i, j}\}$.

\begin{lemma}\label{equivdim}
Consider two integers $i$ and $j$, an ideal $I^\prime$ of $R^\prime$ and
$\phi_{i, j}$ the above application from $R^\prime_{\leqslant i, \leqslant j}$ to
$R_{i,j}$. Then
$$ \dim(R^\prime_{\leqslant i,\leqslant j}/I^\prime_{\leqslant i,\leqslant
j})=\dim(R_{i,j}/\phi_{i, j}(I^\prime_{\leqslant i, \leqslant j})).
$$
\end{lemma}


\begin{demo} 
Remark that $R_{i,j}$, $R^\prime_{\leqslant i, \leqslant j}$, and $I^\prime_{\leqslant i,\leqslant
j}$ and $\phi_{i,j}(I^\prime_{\leqslant i,\leqslant j})$ are finite dimensional
$\Q$-vector spaces.

Moreover, for all $f\in R^\prime_{\leqslant i, \leqslant j}$ and $(i,j)\in \N\times \N$,
$\psi_{i,j}(\phi_{i,j}(f))=f$. Then, $\psi_{i,j}(R_{i,j})=R^\prime_{\leqslant i,
\leqslant j}$ and $\psi_{i,j}(\phi_{i,j}(I^\prime_{\leqslant i, \leqslant j}))=I^\prime_{\leqslant
i, \leqslant j}$. Hence as $\psi_{i,j}$ is an injective morphism, $R_{i,j}$ and $R^\prime_{\leqslant i, \leqslant j}$ on the one hand,
and $I^\prime_{\leqslant i,\leqslant j}$ and $\phi_{i,j}(I^\prime_{\leqslant i,\leqslant j})$ on
the other hand, are isomorphic finite dimensional $\Q$-vector spaces.

Since for vector spaces $E, F$ with $F\subset E$, $\dim(E)=\dim(F)+\dim(E/F)$,
we are done.  \end{demo}

The following lemma is used further. 

\begin{lemma}\label{lemma:phi}
Let $I\subset R$ be a bi-homogeneous ideal such that $X_0$ (resp. $\ell_0$) is
not a zero divisor in $R/I$. Denote by $I^\prime$ the ideal $I+\<X_0-1,
\ell_0-1\>\cap R^\prime$. Then $\phi_{i,j}(I^\prime_{\leqslant i,\leqslant j})=I_{i,j}$.
\end{lemma}

\begin{demo}
Consider a bi-homogeneous polynomial $p\in I_{i,j}$. Obviously,
$\psi_{i,j}(p)\in I^\prime_{\leqslant i, \leqslant j}$. Since the bi-degree of $p$ is, by definition, 
the couple $(i,j)$, $\phi_{i,j}(\psi_{i,j}(p))=p$, this implies $p\in
\phi_{i,j}(I^\prime_{\leqslant i, \leqslant j})$. Thus,
$I_{i,j}\subset\phi_{i,j}(I^\prime_{\leqslant i, \leqslant j})$; we prove now that
$\phi_{i,j}(I^\prime_{\leqslant i, \leqslant j})\subset I_{i,j}$.

{F}rom Lemma~\ref{lemme:bihomgen}, since $I$ is bi-homogeneous, there exists a
finite family of bi-homogeneous polynomials $p_1,\ldots,p_m$ generating
$I$. Consider $p^\prime\in I^\prime_{\leqslant i,\leqslant j}$. Then, there exist
polynomials $q_r$ (for $r\in \{1, \ldots, m\}$), $P$ and $Q$ in $R$, such that
$p^\prime=\sum_{r=1}^{m} q_r. p_r+Q(X_0-1)+P(\ell_0-1)$. Denote by $p$ the
polynomial $\sum_{r=1}^{m}q_r p_r$ and remark that $p\in I$. Since $I$ is
bi-homogeneous, all the bi-homogeneous components of $p$ belong to $I$, and
one just has to consider the case where $p^\prime$ is such that $p$ is
bi-homogenous.

Under this assumption, it is easy to see that $\phi_{i,j}(p^\prime)$ belongs
to $R_{i,j}$ and that there exists a couple $(\alpha, \beta)\in \Z\times \Z$ such
that $\phi_{i,j}(p^\prime)=X_0^{\alpha} . \ell_0^{\beta} . p$. If both
$\alpha,\beta$ are positive, then $\phi_{i,j}(p^\prime)\in I_{i,j}$. If
$\alpha<0$ and $\beta \geqslant 0$, then
$X_0^{-\alpha}\phi_{i,j}(p^\prime)=\ell_0^\beta p\in I$. Suppose
$\phi_{i,j}(p^\prime)\notin I$. This would imply that $X_0$ is a zero-divisor
in $R/I$, which is not possible by assumption. Moreover, since
$\phi_{i,j}(p^\prime)$ has bi-degree $(i, j)$, $\phi_{i,j}(p^\prime)$ belongs
to $I_{i,j}$. Similar arguments allow us to conclude when $\alpha \geqslant 0$ and
$\beta <0$ and when both $\alpha$ and $\beta$ are negative. 
\end{demo}


Given a polynomial $f\in \Q[X_0, \ldots, X_n,\ell_0, \ldots, \ell_k]$, and
$\mathbf{A}\in GL_{n+k+2}(\Q)$, we denote by $f^\mathbf{A}$ the polynomial
obtained by performing the change of variables induced by $\mathbf{A}$ on
$f$. We denote by $I^\mathbf{A}$ the ideal generated by $f_1^\mathbf{A},
\ldots, f_s^\mathbf{A}$. In the sequel, we consider exclusively matrices
$\mathbf{A}$ such that the action of $\mathbf{A}$ on $R$ is a bi-graded
isomorphism of bi-degree $(0,0)$ on $R$ with respect to the variables $X_0,
\ldots, X_n$ and $\ell_0, \ldots, \ell_k$, i.e. for all homogeneous linear
forms $u\in \Q[X_0, \ldots, X_n]$ (resp. $v\in \Q[\ell_0, \ldots, \ell_k]$),
$u^\mathbf{A}$ (resp. $v^\mathbf{A}$) is a homogeneous linear form in
$\Q[X_0, \ldots,  X_n]$ (resp. $\Q[\ell_0, \ldots, \ell_k]$).

\begin{lemma}\label{lemma:3}
Let $I\subset R$ be a bi-homogeneous ideal, then $I$ and $I^\mathbf{A}$ have
the same Hilbert bi-series.
\end{lemma}

\begin{demo} The action of $\mathbf{A}$ on $R$ is an
  isomorphism of bi-graded ring of bi-degree $(0,0)$, the inverse
  action of $\mathbf{A}$ is the action of $\mathbf{A}^{-1}$. Thus, $R_{i,j}$
  equals $R_{i,j}^\mathbf{A}$ and, if $E\subset R$ is a $\Q$-vector
  space, $\dim(E)=\dim(E^\mathbf{A})$.  Since for vector spaces $E, F$
  with $F\subset E$, $\dim(E)=\dim(F)+\dim(E/F)$, we have
  $\dim(R_{i,j}/I_{i,j})=\dim(R_{i,j}^\mathbf{A}/I^\mathbf{A}_{i,j})$
  which implies the equality of the Hilbert bi-series of $I$ and
  $I^\mathbf{A}$.
\end{demo}

\begin{lemma}\label{affbihomseries}
Let $I\subset R$ be a bi-homogeneous ideal which is an intersection of
admissible ideals. Then, for a generic choice of homogeneous linear form $u_1$
(resp. $v_1$) in $\Q[X_0, \ldots, X_n]$ (resp. $\Q[\ell_0, \ldots, \ell_k]$)
the affine Hilbert bi-series of $I+\<u_1-1, v_1-1\>$ equals the Hilbert
bi-series of $I$. 
\end{lemma}

\begin{demo}
{F}rom Remark~\ref{remark:dim2}, since is $u_1$ and $v_1$ are chosen generically
they do not divide $0$ is $R/I$. Choose $\mathbf{A}\in GL_{n+k+2}(\Q)$ such
that the action of $\mathbf{A}$ on $R$ is a bi-graded isomorphism of bi-degree
$(0,0)$ and such that $u_1^\mathbf{A}=X_0$ and
$v_1^\mathbf{A}=\ell_0$. Using Lemma~\ref{lemma:3}, the Hilbert series of $I$
equals the one of $I^\mathbf{A}$. {F}rom Lemma~\ref{equivdim} and
Lemma~\ref{lemma:phi}, the affine Hilbert bi-series of $I^\mathbf{A}+\<X_0-1,
\ell_0-1\>\cap \Q[X_1, \ldots,X_n, \ell_1, \ldots, \ell_k]$ equals the
bi-series of $I^\mathbf{A}$ which equals the one of $I$. 

We prove now $\psi^\prime_{i,j}: R^\mathbf{A}_{\leqslant i, \leqslant j}\rightarrow
{R'}^\mathbf{A}_{\leqslant i, \leqslant j}$ sending $f\in R^\mathbf{A}_{\leqslant i, \leqslant j}$
to $f(1, X_1, \ldots, X_n, \penalty-10000 1, \ell_1, \ldots, \ell_k)$ induces
an isomorphism between ${R'}^\mathbf{A}_{\leqslant i, \leqslant
j}/{I^\prime}^\mathbf{A}_{\leqslant i, \leqslant j}$ and $R^\mathbf{A}_{\leqslant i, \leqslant
j}/(I^\mathbf{A}\penalty-10000+\<X_0-1, \ell_0-1\>_{\leqslant i, \leqslant j})$ which is immediate
using the isomorphism between ${R'}^\mathbf{A}_{\leqslant i, \leqslant
j}/{I^\prime}^\mathbf{A}_{\leqslant i, \leqslant j}$ and
$R^\mathbf{A}_{i,j}/I^\mathbf{A}_{i,j}$.
\end{demo}

The following result is used in the proof of Theorem~\ref{thm:affine}.
\begin{proposition}\label{prop:proj-biproj}
  Let $I$ be a bi-homogeneous ideal of $R$. Then, the Hilbert series of $I$
  equals the series obtained by putting $t_1=t_2$ in the Hilbert bi-series of
  $I$.
\end{proposition}

\begin{demo}
  For all $d\in \N$, denote by $R_d$ the $\Q$-vector space of the {\em homogeneous}
  polynomials of $R$ of total degree $d$. For any couple $(i,j)\in\N\times
  \N$, denote by $R_{i,j}$ the $\Q$-vector space of bi-homogeneous polynomials of
  bi-degree $(i,j)$.  By definition of a bi-homogeneous ideal, if $f$ is a
  polynomial of $I$, then the bi-homogeneous components of $f$ also belong to
  $I$. Hence the morphism from $I_d$ to $\Pi_{i+j=d}I_{i,j}$ sending a
  polynomial onto its bi-homogeneous components is defined for any $f$ of $I$
  and is invertible, i.e. is an isomorphism of $\Q$-vector spaces. This proves that
  $\dim(I_d)=\sum_{i+j=d}\dim(I_{i,j})$. With the same arguments, we show that
  $\dim(R_d)=\sum_{i+j=d}\dim(R_{i,j})$. Consequently
$$\dim(R_d/I_d)=\dim(R_d)-\dim(I_d)=\sum_{i+j=d} \dim(R_{i,j})-\dim(I_{i,j})$$
On the other hand, for all $(i,j)\in \N\times\N$,
$\dim(R_{i,j})-\dim(I_{i,j})=\dim(R_{i,j}/I_{i,j})$.

\end{demo}

\begin{proposition}\label{prop:constanttermseries}
Let $I\subset R$ be a bi-homogeneous ideal and let $J=\cap_{\mathcal{Q}\in
{\rm Adm}(I)}\mathcal{Q}$. Suppose that $J$ has Krull dimension $2$. There
exists $(i_0,j_0)\in \N\times\N$ such that for all {\em $i\geqslant i_0$ and $j\geqslant
j_0$}, $\dim(R_{i,j}/J_{i,j})=\dim(R_{i_0,j_0}/J_{i_0,j_0})$ which equals the
bi-degree of $I$.
\end{proposition}

\begin{demo} {F}rom Remark~\ref{rem:bidegadm}, the bi-degree
of $I$ is the bi-degree of the ideal $J=\cap_{\mathcal{Q}\in {\rm
Adm}(I)}\mathcal{Q}$. {F}rom Proposition~\ref{lemme:genericintersection} and
Definition~\ref{def:bidegre}, the bi-degree of $J$ is the degree of the ideal
$J+\<u-1, v-1\>$ where $u$ (resp. $v$) is a generic homogeneous linear form in
$\Q[X_0, \ldots, X_n]$ (resp. $\Q[\ell_0, \ldots, \ell_k]$).

Consider now $\mathbf{A}\in GL_{n+k+2}(\Q)$ such that $u^\mathbf{A}=X_0$ and
$v^\mathbf{A}=\ell_0$, and such that the canonical action of $\mathbf{A}$ on
$R$ is a bi-graded isomorphism on $R$ of bi-degree $(0, 0)$ with respect to the
variables $X_0, \ldots, X_n$ and $\ell_0, \ldots, \ell_k$. Note that the
bi-degree of $J$ equals the one of $J^\mathbf{A}$ which is the degree of
$J^\mathbf{A}+\<X_0-1, \ell_0-1\>$. In the sequel, denote by
${J^\prime}^\mathbf{A}$ the ideal $\left (J^\mathbf{A}+\<X_0-1, \ell_0-1\>
\right )\cap\Q[X_1, \ldots, X_n, \ell_1, \ldots, \ell_k]$

{F}rom Lemma~\ref{lemma:3},
$\dim(R_{i,j}/J_{i,j})=\dim(R^\mathbf{A}_{i,j}/J^\mathbf{A}_{i,j})$.
Moreover, since by definition of $J^\mathbf{A}$,
$X_0$ (resp. $\ell_0$) is not a zero-divisor in
$R^\mathbf{A}/J^\mathbf{A}$, from Lemma~\ref{lemma:phi},
$\dim(R_{i,j}^\mathbf{A}/J^\mathbf{A}_{i,j})=\dim({R^\prime}^\mathbf{A}_{i,j}/
\phi_{i,j}({J^\prime}^\mathbf{A}_{\leqslant i, \leqslant j}))$.
Finally, from Lemma~\ref{equivdim} $\dim({R^\prime}^\mathbf{A}_{i,j}/
\phi_{i,j}({J^\prime}^\mathbf{A}_{\leqslant i, \leqslant
j}))=\dim({R^\prime}^\mathbf{A}_{\leqslant i,\leqslant
j}/{J^\prime}^\mathbf{A}_{\leqslant i, \leqslant j})$. Thus, one has
$\dim(R_{i,j}/J_{i,j})=\dim({R^\prime}^\mathbf{A}_{\leqslant
i,\leqslant j}/{J^\prime}^\mathbf{A}_{\leqslant i, \leqslant j})$. It
is sufficient to prove there exists $(i_0, j_0)\in \N\times \N$ such
that for all $(i,j)\in \N\times\N$ satisfying $i\geqslant i_0$ and
$j\geqslant j_0$, $\dim({R^\prime}^\mathbf{A}_{\leqslant i,\leqslant
j}/{J^\prime}^\mathbf{A}_{\leqslant i, \leqslant
j})\penalty-10000=\dim({R^\prime}^\mathbf{A}_{\leqslant i_0, \leqslant
j_0}/{J^\prime}^\mathbf{A}_{\leqslant i_0, \leqslant j_0})$ and that
$\dim({R^\prime}^\mathbf{A}_{\leqslant i_0, \leqslant
j_0}/{J^\prime}^\mathbf{A}_{\leqslant i_0, \leqslant j_0})$ equals the
degree of ${J^\prime}^\mathbf{A}$. These are a consequence of
$\dim({J^\prime}^\mathbf{A})=0$ and for $i$, $j$ and $d$ large enough
$\dim({R^\prime}^\mathbf{A}_{\leqslant i,\leqslant
j}/{J^\prime}^\mathbf{A}_{\leqslant i, \leqslant
j})=\dim({R^\prime}^\mathbf{A}_{\leqslant
d}/{J^\prime}^\mathbf{A}_{\leqslant d})=\deg({J^\prime}^\mathbf{A})$,
where ${R^\prime}^\mathbf{A}_{\leqslant d}$
(resp. ${J^\prime}^\mathbf{A}_{\leqslant d}$) denotes the set of
polynomials in ${R^\prime}^\mathbf{A}$ (resp. ${J^\prime}^\mathbf{A}$)
of degree less than or equal to $d$.  
\end{demo}

\subsection{Canonical form of the Hilbert bi-series}\label{sec:canform}


In this paragraph, we provide a canonical form of the Hilbert
bi-series of a bi-homogeneous ideal $I\subset R$ with respect to the
integers $C_{d,e}(I)$ where $(d,e)$ lies in the set of admissible
bi-dimensions of $I$.

\begin{lemma}\label{lemma:suite}
Let $I\subset R$ be a bi-homogeneous ideal. For $i=1, \ldots, s$ let $f_i$ be
a bi-homogene\-ous polynomial in $R$ of bi-degree $(\alpha_i, \beta_i)$. For
$i=1, \ldots, s$, denote by $I_i$ the ideal $I+\<f_1, \ldots, f_i\>$ and
suppose that for all $i\in \{1, \ldots, s-1\}$, $f_{i+1}$ is not a divisor of
zero in $R/I_i$.

Then the Hilbert bi-series of $I_s$ equals $\left
(\Pi_{i=1}^s(1-t_1^{\alpha_i} t_2^{\beta_i})\right )\mathcal{H}(I)$.
\end{lemma}

\begin{demo}
We proceed by induction on $s$. Suppose first $s=1$.

Denote by ${\rm ann}_{R/I}(f_1)$ the annihilator of $f_1$ in $R/I$. The
sequence below
$$ 0 \rightarrow {\rm ann}_{R/I}(f_1)\longrightarrow R/I
{\stackrel{f_1}{\longrightarrow}} R/I \longrightarrow
R/(I+\<f_1\>)\rightarrow 0
$$ is exact. Since $f_1$ is not a zero divisor in $R/I$, ${\rm
ann}_{R/I}(f_1)$ is $0$. Remark that, since $f_1$ is bi-homogeneous,
and ${I}_{i,j}$ and $R_{i,j}$ contain only bi-homogeneous polynomials,
the following one
$$ 0 \longrightarrow R_{i,j}/{I}_{i,j}
{\stackrel{f_1}{\longrightarrow}} R_{i+\alpha_1,j+\beta_1}/{I}_{i+\alpha_1,j+\beta_1} \longrightarrow
R_{i,j}/{(I+\<f_1\>)}_{i,j}\rightarrow 0
$$ is also exact. Thus, classically, the alternate sum of the
dimension of the vector spaces of this exact sequence is null. Adding
these sums for all $(i,j)\in \N\times \N$, one obtains:
$$
(1-t_1^{\alpha_1} t_2^{\beta_1})\mathcal{H}(I)=\mathcal{H}(I_1)
$$ where $(\alpha_1, \beta_1)$ is the bi-degree of $f_1$. Suppose now
the result to be true for $s-1$, i.e. the Hilbert bi-series of
$I_{s-1}$, $\mathcal{H}(I_{s-1})$, equals $\displaystyle\left
(\prod_{i=1}^{s-1}(1-t_1^{\alpha_i}t_2^{\beta_i})\right
)\mathcal{H}(I)$. Since, by assumption, $f_s$ is not a divisor of $0$
in $R/I_{s-1}$, the following sequence is exact:
$$ 0 \rightarrow {\rm ann}_{R/I_{s-1}}(f_s)\longrightarrow R/I_{s-1}
{\stackrel{f_s}{\longrightarrow}} R/I_{s-1} \longrightarrow
R/(I_{s-1}+\<f_s\>)\rightarrow 0
$$ which implies, as above, that
$\mathcal{H}(I_s)=(1-t_1^{\alpha_s}t_2^{\beta_s})\mathcal{H}(I_{s-1})$ and
then, using the induction hypothesis, $\mathcal{H}(I_s)=\left
(\Pi_{i=1}^{s}(1-t_1^{\alpha_i}t_2^{{\beta_i}})\right )\mathcal{H}(I)$ 
\end{demo}

\begin{lemma}\label{lemme:formeR}
The Hilbert bi-series of $R$ is $\frac{1}{( 1 - t_1 )^{n+1} ( 1 - t_2 )^{k+1}}$
\end{lemma}

\begin{demo}
  This is an application of Lemma \ref{lemma:suite}, considering the sequence
  $X_0,\ldots,\penalty-10000X_n,\ell_0,\ldots,\ell_k$, which is a regular one. 
\end{demo}

\begin{proposition}\label{philippe}
Let $I\subset R$ be a bi-homogeneous ideal. Then the Hilbert bi-series of $I$
has the form :
  $$\mathcal{H} ( I ) = \frac{P ( t_1, t_2 )}{( 1 - t_1 )^{n+1} ( 1 - t_2 )^{k+1}} $$
  where $P ( t_1, t_2 )$ is a polynomial of $\Z [ t_1, t_2 ]$.
\end{proposition}

\begin{demo}
We first exhibit a free finite bi-graded resolution of $I$. Next,
considering each bi-homogeneous part of such a resolution, we conclude by
using the alternate summation of the exhibited bi-graded resolution.

{F}rom Lemma~\ref{lemme:bihomgen}, since $I$ is bi-homogeneous, it is
generated by a finite family of bi-homogeneous polynomials $f_1,
\ldots, f_s$, and denote by $(\alpha_i, \beta_i)$ the bi-degree of
$f_i$ for $i=1, \ldots, s$. Thus, $I$ is an $R$-module of finite
type. {F}rom Hilbert's syzygies theorem (see~\cite[p. 208]{perrin}),
$I$ admits a finite free graded resolution :
$$ 0 \longrightarrow L_k=\bigoplus_{i=1}^{d_k}R(\gamma_{k,i})) {\stackrel{\phi_k}{\longrightarrow}} \cdots
  {\stackrel{\phi_2}{\longrightarrow}} L_1=\bigoplus_{i=1}^{d_1}R(\gamma_{1,i})
  {\stackrel{\phi_1}{\longrightarrow}} I \longrightarrow 0 $$
  The quantities $\gamma_i$ are integer shift of the usual graduation
  of $R$ making the morphisms $\phi_i$ homogeneous and of degree $0$.
  The morphism $\phi_1$ from $L_1$ to $I$, sends the element $( p_1,
  \ldots, p_s )$ of $L_1$ to the element $\Sigma_{i = 1}^s p_i f_i$ of
  $I$. 

  In order to make $\phi_1$ a bi-homogeneous morphism of
  bi-degree $(0,0)$, 
  we shift the bi-graduation on each $R$ by the bi-degree of the $f_i$'s : the
  $R$-module $L_1$ becomes $L_1 = \bigoplus_{i = 1}^s R ( (\alpha_i,\beta_i)
  )$ instead of $\bigoplus_{i=1}^s R(\gamma_{1,i})$ (where for $i=1, \ldots,
  s$, $\gamma_{1,i}=\alpha_i+\beta_i$).  

  Considering the system of generators of the syzygies between the $f_i$'s which
  are used in the above resolution, say $r_1, \ldots, r_{d_1} \in L_1$, one
  sees that, as the $f_i$'s are bi-homogeneous, all bi-homogeneous component of
  the $r_j$ are also some syzygies. This means that we can restrict ourselves
  to consider that the $r_i$'s are bi-homogeneous. 

  As above, in order to make $\phi_2$ bi-homogeneous of bi-degree $(0,0)$ we
  shift the bi-graduation of $L_2$ accordingly to the bi-degrees of the $r_j$.
  We apply the same process to the other syzygy modules to finally get a
  bi-graded free resolution of $I$.

  This free bi-graded resolution is an exact sequence, and the morphisms are
  bi-homogeneous of bi-degree $(0,0)$. Thus, the alternate sum of the
  dimensions of the bi-graded parts of degree $( \alpha, \beta )$ of the $L_i$
  and of $I$ is null.

  Remark now that from Lemma \ref{lemme:formeR} the Hilbert bi-series of $R$
  with the bi-graduation shifted of $( \alpha, \beta )$ is
  $\frac{t_1^{\alpha} t_2^{\beta}}{(1-t_1)^{n+1}(1-t_2)^{k+1}}$.

  Consequently, the expression of the Hilbert bi-series of $I$ is obtained by
  summing these Hilbert bi-series, as the denominators of these fractions are
  always the same, the sum can be performed only on the numerators to finally
  get a polynomial $P(t_1,t_2)\in\Z[t_1, t_2]$ such that the Hilbert bi-series
  of $I$ equals $P(t_1, t_2)/(1-t_1)^{n+1}(1-t_2)^{k+1}$. 
\end{demo}

\begin{lemma}\label{new1}
Let $I\subset R$ be a bi-homogeneous ideal and $J=\cap_{\mathcal{Q}\in
{\rm Adm}(I)} \mathcal{Q}$. There exists a couple $(i_0, j_0)\in
\N\times\N$ such that for all $(i,j)\in \N\times\N$ satisfying
$i\geqslant i_0$ and $j\geqslant j_0$:
$$
\dim\left (\frac{R_{i,j}}{I_{i,j}}\right )=\dim\left (\frac{R_{i,j}}{J_{i,j}}\right )
$$
\end{lemma}
\begin{demo}
Denote by $\mathcal{X}$ (resp. $\mathcal{L}$) the intersection of the
isolated primary ideals $\mathcal{Q}$ belonging to a minimal primary
decomposition of $I$ such that there exists $N\in \N$ such that
$\langle X_0, \ldots, X_n\rangle^N\subset \mathcal{Q}$ (resp. $\langle
\ell_0, \ldots, \ell_k\rangle^N\subset \mathcal{Q}$). One has
$I=J\cap\mathcal{X}\cap \mathcal{L}$. We show below that there exists
$(i_0, j_0)\in \N\times \N$ such that for all $i\geqslant i_0$ and all
$j\geqslant j_0$, $I_{i,j}=J_{i,j}$.

Let $i_0$ be the smallest integer such that $\langle X_0, \ldots,
X_n\rangle^{i_0}\subset \mathcal{X}$, (resp. $j_0$ the smallest integer
such that $\langle \ell_0, \ldots, \ell_k\rangle^{j_0}\subset
\mathcal{L}$). Then for all $i\geqslant i_0$ and all $j\in\N$,
$\mathcal{X}_{i,j}=R_{i,j}$. Similarly, for all $j\geqslant j_0$ and all
$i\in\N$, $\mathcal{L}_{i,j}=R_{i,j}$.

Remark that for all $(i,j)\in\N\times\N$, $I_{i,j}=\left (J\cap
\mathcal{X}\cap\mathcal{L}\right )_{i,j}$ equals
$J_{i,j}\cap\mathcal{X}_{i,j}\cap\mathcal{L}_{i,j}$. Then, for all
$i\geqslant i_0$ and all $j\geqslant j_0$, $I_{i,j}$ equals
$J_{i,j}$. This allows to conclude.  
\end{demo}

\begin{lemma}\label{new2}
Let $I\subset R$ be a bi-homogeneous ideal, $J=\cap_{\mathcal{Q}\in
{\rm Adm}(I)}\mathcal{Q}$ and $f\in R$ be a bi-homogeneous polynomial
of bi-degree $(\alpha, \beta)$ which does not divide $0$ in $R/J$.
Denote by $\sum_{i,j} a_{i,j}t_1^i t_2^j$ the Hilbert bi-series of
$I+\<f\>$ and by $\sum_{i,j} b_{i,j} t_1^i t_2^j$ the Hilbert
bi-series of $I$ times $(1-t_1^\alpha t_2^\beta)$.

There exists a couple $(i_0, j_0)\in \N\times \N$ such that for all
$(i,j)\in \N\times \N$ satisfying $i\geqslant i_0$ and $j\geqslant
j_0$, $a_{i,j}=b_{i,j}$.
\end{lemma}

\begin{demo}
We denote by $\sum_{i,j}c_{i,j} t_1^i t_2^j$ the Hilbert bi-series of
$J+\<f\>$ and by $\sum_{i,j}d_{i,j} t_1^i t_2^j$ the Hilbert bi-series
of $J$ times $(1-t_1^\alpha t_2^\beta)$.  {F}rom Lemma
\ref{lemma:suite}, $H(J+\<f\>)$ equals $(1-t_1^\alpha
t_2^\beta)H(J)$. This implies that for all $(i,j)\in \N\times\N$ $d_{i,j}=c_{i,j}$. 
In the sequel, we prove that there exists a couple $(i_0, j_0)\in
\N\times\N$ such that for all $(i,j)\in \N\times\N$ satisfying
$i\geqslant i_0$ and $j\geqslant j_0$, $c_{i,j}=a_{i,j}$ and
$b_{i,j}=d_{i,j}$.

Remark now that ${\rm Adm}(J+\<f\>)={\rm Adm}(I+\<f\>)$. Thus, from
Lemma~\ref{new1}, there exists a couple $(i_1, j_1)\in \N\times\N$ such
that for $(i,j)\in \N\times\N$ satisfying $i\geqslant i_1$ and
$j\geqslant j_1$, $\dim\left (\frac{R_{i,j}}{(\cap_{\mathcal{Q}\in
{\rm Adm}(I+\<f\>)}\mathcal{Q})_{i,j}}\right
)=c_{i,j}=a_{i,j}$. Similarly, applying Lemma~\ref{new1} to $I$ and $J$
implies the existence of a couple $(i_2, j_2)\in \N\times\N$ such that
for all $(i,j)$ satisfying $i\geqslant i_2$ and $j\geqslant j_2$,
$b_{i,j}=d_{i,j}$. Choosing $i_0=\max(i_1, i_2)$ and $j_0=\max(j_1,
j_2)$ allows us to conclude.

\end{demo}

The following result provides a canonical form of the Hilbert
bi-series of a bi-homogeneous ideal. It generalizes the results
of~\cite{Rem01a} which yields a similar result in the case of a prime
admissible bi-homogeneous ideal.

\begin{proposition}\label{prop:formecanonique}
Let $I\subset R$ be a bi-homogeneous ideal, $D$ be the maximal
Krull-dimension of its isolated admissible primary components and
$\mathcal{D}=\{(d, e)\in \N\times\N\mid d+e+2=D\}$. For $(i,j)\in
\N\times \N$ such that $i+j\leqslant D-3$, there exist $c_{i,j}\in \Z$
and a polynomial $Q\in \Z[t_1, t_2]$ such that the Hilbert bi-series
of $I$ equals:
{\small$$ \left (\sum_{(d,e)\in \mathcal{D}}
\frac{C_{d,e}(I)}{(1-t_1)^{d+1}(1-t_2)^{e+1}}\right )+\left
(\sum_{-1\leqslant i+j\leqslant D-3}\frac{c_{i,j}}{(1-t_1)^{i+1}
(1-t_2)^{j+1}}\right )+Q(t_1, t_2)
$$}
\end{proposition}

\begin{demo}
{F}rom Proposition~\ref{philippe}, there exists a polynomial $P\in \Z[t_1,
t_2]$ such that the Hilbert bi-series of $I$ we denote by $\mathcal{H}(I)$ equals:
$$
\frac{P(t_1, t_2)}{(1-t_1)^{n+1}(1-t_2)^{k+1}}
$$ Writing the polynomial $P$ on the basis $\{(1-t_1)^i (1-t_2)^j\mid (i,j)\in
\N\times\N\}$, one obtains the existence of a polynomial $Q\in \Z[t_1, t_2]$
and of integers $c_{i,j}\in \Z$ such that the Hilbert bi-series of $I$ equals:
\begin{equation}\label{eqtoto}
\sum_{0\leqslant i+j\leqslant n+k+2} \frac{c_{i,j}}{(1-t_1)^i (1-t_2)^j}+Q.
\end{equation}
Given a couple $(d,e)\in \mathcal{D}$, for a generic choice of
homogeneous linear forms $u_1, \ldots, u_d$ (resp. $v_1, \ldots, v_e$) in
$\Q[X_0, \ldots, X_n]$ (resp. $\Q[\ell_0, \ldots, \ell_k]$), one has:
\begin{itemize}
  \item {f}rom Proposition~\ref{lemme:genericintersection}, $I^{(d,
e)}=I+\<u_1, \ldots, u_d, v_1, \ldots, v_e\>$ is a bi-homogeneous ideal of
bi-dimension $(0,0)$, 
  \item from Remark \ref{remark:dim2}, $u_i$ (resp. $v_j$) does not divide zero in
  $\frac{R}{\left (\cap_{\mathcal{Q}\in {\rm Adm}(I+\<u_1, \ldots,
  u_{i-1}\>)}\mathcal{Q}\right )}$, 
  \item from Proposition \ref{prop:constanttermseries} and
  Lemma~\ref{new1}, there exists $(i_0, j_0)\in \N\times \N$ such that for
  all $(i,j)\in \N\times \N$ such that $i\geqslant i_0$ and
  $j\geqslant j_0$, the term of index $(i,j)$ in the Hilbert bi-series
  of $I^{(d,e)}$ equals the bi-degree of $I^{(d,e)}$,
  \item from Lemma \ref{lemma:suite} and Lemma~\ref{new1}, there exists $(i_1,
  j_1)\in \N\times \N$ such that for all $(i,j)\in \N\times \N$ such
  that $i\geqslant i_1$ and $j\geqslant j_1$, the term of index
  $(i,j)$ in the Hilbert bi-series of $I^{(d,e)}$ equals the term of
  index $(i,j)$ in the bi-series $(1-t_1)^d(1-t_2)^e\mathcal{H}(I)$.
\end{itemize}
Remark now that there exists a polynomial $\widetilde{Q}\in \Z[t_1, t_2]$ such
that:
\begin{eqnarray*}
(1-t_1)^d(1-t_2)^e\mathcal{H}(I) &=
 &\sum_{i,j}\frac{c_{i,j}}{(1-t_2)^{i-d}(1-t_2)^{j-e}} +(1-t_1)^d(1-t_2)^e Q \\
 & = & \frac{c_{d+1, e+1}}{(1-t_1)(1-t_2)} +
 \sum_{\begin{subarray}{c}i\geqslant d, \\j\geqslant
 e\end{subarray}}\frac{c_{i,j}}{(1-t_1)^{i-d}(1-t_2)^{j-e}} +\widetilde{Q}
\end{eqnarray*}
which implies that $c_{d+1, e+1}=C_{d,e}(I)$ and $c_{i,j}=0$ if $i\geqslant d$
and $j\leqslant e$.

\end{demo}

\subsection{Properties of the bi-degree of a bi-homogeneous ideal}\label{sec:ptesbideg}

In this paragraph, we study the bi-degree of $I+\<f\>$ when $I$ (resp. $f$)
is a bi-homogeneous ideal (resp. polynomial) of $R$ and exhibit how it is
related to the one of $I$.

The following result appears in a slightly different form
in~\cite{VdW29} and~\cite[Lemma 2.11]{Rem01a} in the case of a prime
ideal.

\begin{proposition}\label{prop3}
Let $I\subset R$ be a bi-homogeneous ideal, $J=\cap_{\mathcal{Q}\in {\rm
Adm}(I)}\mathcal{Q}$, $D$ be the dimension of $J$, and $f\in R$ a
non-divisor of zero in $R/J$. Suppose that $J$ is equidimensional. 

Then, the bi-degree of $I+\<f\>$ is equal to:
$$
\begin{array}{r}
\alpha\left (\sum_{(d, e)\mid d+e+3=D} C_{d+1, e}(I)\right )+\beta\left
    (\sum_{(d, e)\mid d+e+3=D} C_{d, e+1}(I)\right )
\end{array}
$$
\end{proposition}

\begin{demo} 
{F}rom Proposition~\ref{prop:formecanonique}, the Hilbert bi-series of $J$ can
be written as:
$$ \sum_{d+e+2=D}\frac{C_{d,e}(J)}{(1-t_1)^{d+1}(1-t_2)^{e+1}}+
\sum_{i+j+2\leqslant D-1}\frac{c_{i,j}}{(1-t_1)^{i+1}(1-t_2)^{j+1}} + Q(t_1, t_2)
$$ {F}rom Lemma~\ref{lemma:suite}, since $f$ does not divide zero in $R/J$,
the Hilbert bi-series of $J+\<f\>$ equals the Hilbert bi-series of $J$
multiplied by $(1-t_1^\alpha t_2^\beta)$. Since
\begin{eqnarray*}
1-t_1^\alpha t_2^\beta & = &1-t_1^\alpha + t_1^\alpha (1-t_2)^\beta \\
& = & (1-t_1)\sum_{p=0}^{\alpha-1}t_1^p + t_1^\alpha (1-t_2)
\sum_{q=0}^{\beta-1} t_2^q
\end{eqnarray*}
some easy computations show that the Hilbert bi-series of $J+\<f\>$ has
the form:
\begin{eqnarray*}
 \sum_{d+e+2=D-1} &\frac{\alpha C_{d+1,e}(J)+\beta C_{d,
e+1}(J)}{(1-t_1)^{d+1}(1-t_2)^{e+1}} +\\ &\sum_{i+j+2\leqslant
D-2}\frac{\widetilde{c}_{i,j}}{(1-t_1)^{i+1}(1-t_2)^{j+1}} +
\widetilde{Q}(t_1, t_2)
\end{eqnarray*}
for some $\widetilde{c}_{i,j}\in \Z$ and $Q\in \Z[t_1, t_2]$.  Thus,
from Proposition~\ref{prop:formecanonique}, the bi-degree of
$J+\<f\>$, which equals the bi-degree of $I+\<f\>$ is
$$
\alpha \sum_{d+e+2=D} C_{d+1, e}(I) +\beta\sum_{d+e+2=D} C_{d, e+1}(I).
$$

\end{demo}

The following lemma extends to the bi-projective case a result
of~\cite{Heintz83}.

\begin{lemma}\label{aecrire}
  Let $I$ be a bi-homogeneous ideal, and denote by $J$ the ideal,
  $J=\cap_{\mathcal{Q}\in {\rm Adm}(I)}\mathcal{Q}$.  Suppose that $J$
  is equidimensional and $\dim(J)\geqslant 3$. Let $f$ be a
  bi-homogeneous polynomial of bi-degree $(\alpha,\beta)$ dividing
  zero in $R/J$ and such that $\dim(J+\<f\>)=\dim(J)$. Then, the
  bi-degree of $I+\<f\>$ is less than or equal to the bi-degree of
  $I$.

  Suppose now that there exists $\tilde{f}$ a bi-homogeneous
  polynomial of bi-degree $(\alpha,\beta)$ which does not divide $0$
  in $R/J$. Then, denoting by $D$ the dimension of $J$:
  \begin{itemize}
  \item if $\alpha\neq 0$ and $\beta\neq 0$, $\bideg(I+\<f\>)\leqslant
  \bideg(I+\<\tilde{f}\>)$;
  \item if $\alpha=0$ and $C_{0, D-2}(J)=0$, $\bideg(I+\<f\>)\leqslant
  \bideg(I+\<\tilde{f}\>)$;
  \item if $\beta=0$ and $C_{D-2, 0}(J)=0$, $\bideg(I+\<f\>)\leqslant
  \bideg(I+\<\tilde{f}\>)$.
  \end{itemize}
\end{lemma}

\begin{demo} 
  Let $\mathcal{D}\subset \N\times\N$ be the set of admissible dimensions of
  $J$. Since $\dim(J+\<f\>)=\dim(J)$, $\bideg(J+\<f\>)=\sum_{(d,e)\in
  \mathcal{D} } C_{d,e}(J+\<f\>)$. {F}rom Proposition
  \ref{lemme:genericintersection} there exist $\mathcal{H}_1$ and
  $\mathcal{H}_2$, two Zariski closed subset of
  $(\mathbb{C}^{n+1})^{d+1}\times(\mathbb{C}^{k+1})^{e+1}$ such that if we
  choose $u_1,\ldots,u_d,u_{d+1}$, $v_1,\ldots,v_e,v_{e+1}$ outside
  $\mathcal{H}_1\cup\mathcal{H}_2$, $J+\<(u_1-1),\ldots,u_d,u_{d+1}
  ,(v_1-1),\ldots,v_e,v_{e+1}\>$ and
  $J+\<(u_1-1),\ldots,u_d,u_{d+1},v_1,\ldots,v_e,\penalty-10000
  v_{e+1}\>+\<f\>$ are zero-dimensional ideals.  Hence as
  $J+\<(u_1-1),\ldots,u_d,u_{d+1},(v_1-1),\ldots,v_e,v_{e+1}\>$ is included in
  $J+\<(u_1-1),\ldots,u_d,u_{d+1},(v_1-1),\ldots,v_e,\penalty-10000
  v_{e+1},f\>$,
  $\deg(J+\<(u_1-1),\ldots,u_d,u_{d+1},(v_1-1),\ldots,v_e,v_{e+1}\>+\<f\>)\leqslant
  \deg(J+\<u_1,\ldots,u_d,v_1,\ldots,v_e\>)$. This proves that
  $C_{d,e}(I+\<f\>)=C_{d,e}(J+\<f\>)\leqslant C_{d,e}(J)=C_{d,e}(I)$. Summing these
  equality for all admissible bi-dimension one obtains that
  $\bideg(I+\<f\>)\leqslant \bideg(I)$ which proves the first part of the result.

  Consider now $\widetilde{f}$ a bi-homogeneous polynomial of bidegree
  $(\alpha, \beta)$ which does not divide zero in $R/J$. {F}rom
  the formula given by Proposition \ref{prop3}, the bi-degree of $J+\<\widetilde{f}\>$ is
  obviously greater than the one of $J$ if 
  \begin{itemize}
    \item $\alpha\neq 0$ and $\beta\neq 0$,
    \item or $\alpha=0$ and $C_{D-2, 0}(J)=0$ where $D=\dim(J)$,
    \item or $\beta=0$ and $C_{0,D-2}(J)=0$ where $D=\dim(J)$,
  \end{itemize}
which ends the proof.

\end{demo}

\subsection{Proofs of Theorems~$1$ and $2$}\label{sec:prfthms}



We can now prove Theorem $1$, which we now restate. 

{\bf Theorem $1$} {\em Let $s\in \{1, \ldots, n+k\}$ and $f_1,\ldots,f_s$ be bi-homogeneous
polynomials in $R$ of respective bi-degree $(\alpha_i,\beta_i)$ generating a
bi-homogeneous ideal $I$.  Suppose that there exist at most $n$ $f_i$ such
that $\beta_i=0$ and at most $k$ $f_i$ such that $\alpha_i=0$, then the sum of
the bi-degrees of the bi-homogeneous associated primes of $I$ is bounded by
$\displaystyle\mathcal{B}(f_1,\ldots,f_s)=\sum_{\mathcal{I}, \mathcal{J}}
\left (\Pi_{i\in\mathcal{I}}\alpha_{i}\right ).\left
(\Pi_{j\in\mathcal{J}}\beta_{j}\right ) $ where $\mathcal{I}$ and
$\mathcal{J}$ are disjoint subsets for which the union is $\{1, \ldots, s\}$
such that the cardinality of $\mathcal{I}$ (resp. $\mathcal{J}$) is bounded by
$n$ (resp. $k$).}

{\bf Proof of Theorem~\ref{bezout}.}  
For $i\in \{1, \ldots, s\}$ we denote by $I_i$ the ideal generated by $\langle
f_1, \ldots, f_i\rangle$. Given an ideal $I$, we denote by $\Ass(I)$ the set
of primes associated to $I$. Given $(d,e)\in \N\times\N $, we identify the
cartesian product of the set of $d$ linear homogeneous forms in $\Q[X_0,
\ldots, X_n]$ and $e$ linear homogeneous forms in $\Q[\ell_0, \ldots, \ell_k]$
to $\Q^{d(n+1)}\times\Q^{e(k+1)}$.

Let $\Pf(i)$ be the property: for every couple $(d, e)\in \N\times\N$ such
that $d+e=n+k-i$, there exists a Zariski-closed subset
$\mathcal{H}\subsetneq\Q^{d(n+1)}\times \Q^{e(k+1)}$ such that for every
choice of 
\begin{itemize}
  \item $d$ linear homogeneous forms $u_1, \ldots, u_d$ in $\Q[X_0, \ldots,
X_n]$ generating the ideal denoted by $\mathcal{U}_d$
  \item $e$ linear homogeneous forms $v_1, \ldots, v_e$ in $\Q[\ell_0, \ldots,
\ell_k]$ generating the ideal denoted by $\mathcal{V}_e$
  \item such that $(u_1, \ldots, u_d, v_1, \ldots, v_e)\in
  \left (\Q^{d(n+1)}\times\Q^{e(k+1)}\right )\setminus \mathcal{H}$
\end{itemize}
One has :
$$ \sum_{\Pc\in \Ass(\sqrt{I_i})}\bideg\left (\Pc + \mathcal{U}_d+
\mathcal{V}_e \right )\leq \sum_{\begin{subarray}{c}|\mathcal{A}|=n-d,|\mathcal{B}|=k-e\\
\mathcal{A}\cap\mathcal{B}=\emptyset,\mathcal{A}\cup\mathcal{B}=\{1, \ldots,
i\}\end{subarray}} \left (\prod_{p\in \mathcal{A}} \alpha_p\prod_{q\in \mathcal{B}}
\beta_q\right )
$$ 

By definition~\ref{def:bidegre} of the bi-degree of a bi-homogeneous ideal
$\Pf(1)$ is true. Let us show now that $\Pf(i)$ implies $\Pf(i+1)$


Remark that $$(\bigcap_{\mathcal{P}\in \Ass(\sqrt{I_i})}\Pc)+\<f_{i+1}\>\subset
\bigcap_{\Pc\in \Ass(\sqrt{I_i})}(\Pc+\<f_{i+1}\>)\subset \bigcap_{\Pc\in
\Ass(\sqrt{I_i})}\left (\sqrt{\Pc+\<f_{i+1}\>}\right ).$$ Note also that all the
above ideals have the same radical which is $\sqrt{I_{i+1}}$.  Let
$\mathcal{Q}_0\in \Ass(\sqrt{I_{i+1}})$ be a bihomogeneous admissible prime ideal. We
show now that there exists $\mathcal{P}_0\in \Ass(\sqrt{I_i})$ such that
$\mathcal{Q}_0\in \Ass(\sqrt{\mathcal{P}+\<f_{i+1}\>})$.

Notice that the ideal $J=\bigcap_{\Pc\in\Ass(\sqrt{I_i})}
\sqrt{\Pc+\<f_{i+1}\>}$ equals the ideal $\sqrt{I_{i+1}}=\cap_{\Qc\in\Ass(\sqrt{I_{i+1}})}\Qc$.
Consider 
$h\in \bigcap_{\Qc\in \Ass(\sqrt{I_i})\setminus \{\Qc_0\}}\Qc$.
The saturation of $J$ by $h$, $J:h^\infty=\{f\in R\mid \exists p \in
\N \mid h^p f\in J\}$ equals $\Qc_0$ ($\Qc_0$ is prime) which implies that
$\bigcap_{\Pc\in\Ass(\sqrt{I_i})} \left
(\sqrt{\Pc+\<f_{i+1}\>}:h^\infty\right )=\Qc_0$. For all $\Pc\in
\Ass(\sqrt{I_i})$, $\sqrt{\Pc+\<f_{i+1}\>}:h^\infty$ is the
intersection of the ideals in $\Ass(\sqrt{\Pc+\<f_{i+1}\>})$ which do
not contain $h$. Thus, there exists $\Pc\in \Ass(\sqrt{I_i})$ such
that $\sqrt{\Pc+\<f_{i+1}\>}:h^\infty= \Qc_0$ which implies that
$\Qc_0\in \Ass(\sqrt{\Pc+\<f_{i+1}\>})$.




Consider a prime ideal $\Pc$ of $\Ass(\sqrt{I_i})$, let $(d,e)\in
\N\times\N$ be such that $d+e=n+k-i-1$, and let $u_1,\ldots,u_{d}$
(resp. $v_1,\ldots,v_{e}$) be linear homogeneous forms in $\Q[X_0,
\ldots, X_n]$ (resp. in $\Q[\ell_0, \ldots, \ell_k]$), then denoting
by $\Uc_d$ the ideal generated by $u_1,\ldots,u_{d}$ (resp. $\Vc_{e}$
the one generated by $v_1,\ldots,v_{e}$), one has :
\begin{eqnarray*}
\Pc+\<f_{i+1}\>+\Uc_{d}+\Vc_{e}&\subset
\sqrt{\Pc+\<f_{i+1}\>}+\Uc_{d}+\Vc_{e}&\subset \\(\bigcap_{\Qc_{}\in
\Ass(\Pc+\<f_{i+1}\>)} \Qc_{})+\Uc_{d}+\Vc_{e}&\subset
\bigcap_{\Qc_{}\in \Ass(\sqrt{\Pc+\<f_{i+1}\>})}
(\Qc_{}+\Uc_{d}+\Vc_{e})
\end{eqnarray*}

Since $\Pc$ is prime, $\Pc+\<f_{i+1}\>$ is equidimensionnal which
implies that for all $\mathcal{Q}\in \Ass(\sqrt{\Pc+\<f_{i+1}\>})$,
$\dim(\mathcal{Q})=\dim(\Pc+\<f_{i+1}\>)$. 

So from the above four inclusions, one deduces that:
$$ \bideg(\Pc+\<f_{i+1}\>+\Uc_{d}+\Vc_{e})\geq\sum_{\mathcal{R}\in
\Ass(\sqrt{\Pc+\<f_{i+1}\>})} \bideg(\mathcal{R}+\Uc_{d}+\Vc_{e})
$$

As shown above, for all $\Qc\in\Ass(\sqrt{I_{i+1}})$ there exists at least
one ideal $\Pc\in\Ass(\sqrt{I_i})$ such that $\Qc\in\Ass(\sqrt{\Pc+\<f_{i+1}\>})$. Thus, the
above inequality implies that:
$$ \sum_{\Qc\in \Ass(\sqrt{I_{i+1}})}\bideg\left (\Qc + \mathcal{U}_d+
\mathcal{V}_e \right )\leq \sum_{\mathcal{P}\in
\Ass(\sqrt{I_{i}})}\bideg\left (\mathcal{P} +\<f_{i+1}\> +
\mathcal{U}_d+ \mathcal{V}_e \right )
$$

{F}rom the theorem's assumptions, $I_{i+1}$ is generated by $f_1, \ldots,
f_{i+1}$, with $i\leq n+k-1$. Thus, since the couple $(d,e)$ is such
that $d+e=n+k-i-1$, for all $\mathcal{P}\in \Ass(\sqrt{I_i})$, one has
$\dim(\Pc+\mathcal{U}_d+\mathcal{V}_e)\geq 2$ which allows us to use
Lemma \ref{aecrire} and Proposition \ref{prop3} to prove the existence
of a Zariski-closed subset $\mathcal{A}\subsetneq
\C^{n+1}\times\C^{k+1}$ such that if $u$ and $v$ are homogeneous
linear forms of $\Q[X_0, \ldots, X_n]$ and $\Q[\ell_0, \ldots,
\ell_k]$ and $(u,v)$ is chosen outside the Zariski-closed subset
$\mathcal{A}$
$$\sum_{\mathcal{\mathcal{Q}}\in \Ass(\sqrt{I_{i+1}})}\bideg\left
(\mathcal{Q} + \mathcal{U}_d+ \mathcal{V}_e \right )$$ is bounded by
the sum : {\small$$\alpha_{i+1}\sum_{\mathcal{P}\in
\Ass(\sqrt{I_{i}})}\bideg\left (\mathcal{P} +\<u\> + \mathcal{U}_d+
\mathcal{V}_e \right )+\beta_{i+1}\sum_{\mathcal{P}\in
\Ass(\sqrt{I_{i}})}\bideg\left (\mathcal{P} +\<v\> + \mathcal{U}_d+
\mathcal{V}_e \right )$$}

Then, using $\Pf(i)$ on the preceding formulation allows us to state
$\Pf(i+1)$.

Thus, under the assumptions of Theorem~$1$, for $i< s$, the
property $\Pf(i)$ implies $\Pf(i+1)$ and $\Pf(1)$ is true.

We now bound the strong bi-degree of $I$ using $\Pf(s)$.  Let $(d,e)$
be a couple of integers such that $d+e\leq n+k$, and
$\mathcal{H}\subsetneq\C^{d(n+1)}\times\C^{e(k+1)}$ be a
Zarikisi-closed subset such that chosing homogeneous linear forms
$u_1, \ldots, u_d$ (resp. $v_1, \ldots, v_e$) in $\Q[X_0, \ldots,
X_n]$ (resp. $\Q[\ell_0, \ldots, \ell_k]$) outisde $\mathcal{H}$ and
denoting by $\Uc_d$ (resp. $\Vc_e$) be the ideal generated by $u_1,
\ldots, u_d$ (resp. $v_1, \ldots, v_e$), by definition of the
bi-degree, for any prime ideal $\Pc$ such that $\dim(\Pc)>d+e$, one
has :
$$\bideg(\Pc+\Uc_d+\Vc_e)=\sum_{\begin{subarray}{c}(d',e')\in \N\times \N\\
  d+e+d'+e'=\dim(\Pc)-2\end{subarray}}C_{d+d',e+e'}(\Pc).$$

We end the proof looking at the hilbert bi-series of these ideals and remarking that for any associated prime $\Pc$ of
$\sqrt{I}$
$$\bideg(\Pc)\leq\sum_{\begin{subarray}{c}(d,e)\in\N\times
\N\\d+e=n+k-s\end{subarray}}\bideg(\Pc+\Uc_d+\Vc_e).$$ Hence $\Pf(s)$ allows
us to bound the strong bi-degree of $I$ by $\Bc(f_1,\ldots,f_s)$.


We prove now Theorem~\ref{thm:affine} which we now restate:

{\em Consider the mapping :
$$
\phi\; :\; \begin{array}[t]{ccl}
    \Q[X_1, \ldots, X_n, \ell_1, \ldots, \ell_k]&\rightarrow &  \Q[X_0,X_1, \ldots, X_n, \ell_0,\ell_1, \ldots, \ell_k]\\
      f&\mapsto& X_0^{\deg_X(f)}\ell_0^{\deg_\ell(f)} f(\frac{X_1}{X_0}, \ldots,
      \frac{X_n}{X_0}, \frac{\ell_1}{\ell_0}, \ldots,
      \frac{\ell_k}{\ell_0}) \\
  \end{array}
$$ where $\deg_X(f)$ (resp. $\deg_\ell(f)$) denotes the degree of $f$
seen as a polynomial in $\Q(\ell_1, \ldots, \ell_k)[X_1, \ldots, X_n]$
(resp. $\Q(X_1, \ldots, X_n)[\ell_1, \ldots, \ell_k]$). 

Given an ideal $I\subset\Q[X_1, \ldots, X_n, \ell_1, \ldots, \ell_k]$,
denote by $\phi(I)$ the ideal generated by $\{\phi(f)\mid f\in
I\}\subset \Q[X_0, \ldots, X_n, \ell_0, \ldots, \ell_k]$.

Then, $\phi(I)$ is a bi-homogeneous ideal and the sum of the degrees
of the isolated primary components of $I$ is bounded by the strong
bi-degree of $\phi(I)$. }

{\bf Proof of Theorem~\ref{thm:affine}.}
The first assertion is obvious. We focus on the second one. Denote by $\psi$ the mapping: 
$$ \psi\; :\; \begin{array}[t]{ccl} \Q[X_1, \ldots, X_n, \ell_1,
\ldots, \ell_k]&\rightarrow & \Q[X_0,X_1, \ldots, X_n, \ell_1, \ldots,
\ell_k]\\ f&\mapsto& X_0^{\deg(f)}f(\frac{X_1}{X_0}, \ldots,
\frac{X_n}{X_0}, \frac{\ell_1}{X_0}, \ldots, \frac{\ell_k}{X_0}) \\
  \end{array}
$$ Given an ideal $I\subset \Q[X_1, \ldots, X_n, \ell_1, \ldots,
\ell_k]$, we denote by $\psi(I)$ the homogeneous ideal generated by
$\{\psi(f)\mid f\in I\}\subset \Q[X_0,X_1, \ldots, X_n, \ell_1,
\ldots, \ell_k]$.

Following~\cite{CLO}, if $\mathcal{Q}_1, \ldots, \mathcal{Q}_r$ are the
primary ideals of a minimal primary decomposition of $I$,
$\phi(I)=\cap_{i=1}^r \phi(\mathcal{Q}_i)$ and $\psi(I)=\cap_{i=1}^r
\psi(\mathcal{Q}_i)$.

Thus, if $I$ is equidimensional, $\phi(I)$ and $\psi(I)$ are
equidimensional. Additionnally, if $I$ is radical, $\phi(I)$ and
$\psi(I)$ are radical.

Suppose now that $I$ is an equidimensional ideal.  Given a
polynomial $f\in \Q[X_1, \ldots, X_n, \ell_1, \ldots, \ell_k]$, remark
that $\phi(f)$ divided by $X_0-\ell_0$ with respect to $\ell_0$ equals
$$X_0^{\min(\deg_X(f), \deg_\ell(f))}\psi(f).$$ This proves that the
ideal $\left (\phi(I)+\langle X_0-\ell_0\rangle\right )\cap \Q[X_0,
\ldots, X_n, \ell_1, \ldots, \ell_k]$ saturated by $X_0$, denoted by
$J$ in the sequel, equals $\psi(I)$. This implies that the degree of
$J$ equals the one of $\psi(I)$. Thus the degree of $\psi(I)$ is
bounded by the degree of $\left (\phi(I)+\langle
X_0-\ell_0\rangle\right )\cap \Q[X_0, \ldots, X_n, \ell_1, \ldots,
\ell_k]$ which is itsself bounded by the degree of $\phi(I)+\langle
X_0-\ell_0\rangle$. {F}rom B\'ezout's theorem (see \cite[Fulton]{}),
the sum of the degrees of the isolated primary components of
$\phi(I)+\langle X_0-\ell_0\rangle$ is bounded by the degree of
$\phi(I)$, since $\phi(I)$ is equidimensional.

Thus, the degree of $\psi(I)$ is bounded by the one of $\phi(I)$.

On the one hand, the degree of $I$ equals the one of $\psi(I)$ (see
\cite[a expliciter]{CLO}). On the other hand, from
Proposition~\ref{prop:formecanonique} and
Proposition~\ref{prop:proj-biproj}, the degree of $\phi(I)$ equals the
bi-degree of $\phi(I)$. Finally, if $I$ is equidimensional, its degree
is bounded by the bi-degree of $\phi(I)$.

If $I$ is not equidimensional, it is sufficient to apply the above to
each isolated primary component of $I$ since if $\mathcal{Q}_1,
\ldots, \mathcal{Q}_r$ are the primary ideals of a minimal primary
decomposition of $I$, $\psi(I)=\cap_{i=1}^r \psi(\mathcal{Q}_i)$ and
$\phi(I)=\cap_{i=1}^r \phi(\mathcal{Q}_i)$.

\begin{corollary}\label{cor:final}
  Let $S$ be a finite polynomial family in $\Q[X_1, \ldots, X_n, \ell_1,
  \ldots,\ell_k]$ and $I$ be the ideal generated by $S$ which is supposed to
  be radical. Consider the ideal $J$ of $\Q[X_0, \ldots, X_n, \ell_0, \ldots,
  \ell_k]$ generated by $\{ \phi(f)\mid f\in S\}$.

  Then, the sum of the degrees of the irreducible components of $I$ is
  bounded by the strong bi-degree of $\sqrt{J}$.
\end{corollary}

\begin{demo}
{F}rom Theorem~\ref{thm:affine}, the degree of $I$ is bounded by the
strong bi-degree of $\phi(I)$. Thus, it is sufficient to prove that
the strong bi-degree of $\phi(I)$ is bounded by the one of $\sqrt{J}$.


Since, from~\cite{CLO}, for all $f\in \phi(I)$, there exists $p$ and
$q$ in $\N$ such that $X_0^p\ell_0^q f$ belongs to $J$, $\phi(I)$
equals the ideal obtained by saturating $J$ by $X_0$ and
$\ell_0$. Since $I$ is radical, following the proof of
Theorem~\ref{thm:affine}, $\phi(I)$ is radical. Then, each prime
component $\mathcal{Q}$ of $\phi(I)$ is an isolated primary component
of $J$.
Hence is a prime component of $\sqrt{J}$.
This implies that the strong bi-degree of $\sqrt{J}$ (being the sum of
the degree of the prime components of $\sqrt{J}$) bounds the one of
$\phi(I)$ as $\phi(I)$ is radical and as all the prime components of
$\phi(I)$ can be found among the ones of $\sqrt{J}$.
\hfill
\end{demo}

\section{Degree bounds on the critical locus of a projection}
\label{sec:crit}

Consider a polynomial family $(f_1, \ldots, f_s)$ in $\Q[X_1, \ldots,
X_n]$ generating a radical ideal such that the algebraic variety
$\mathcal{V}\subset \C^n$ defined by $f_1=\cdots=f_s=0$ is smooth,
$f_{s+1}\in \Q[X_1, \ldots, X_n]$ and the polynomial mapping
$\widetilde{f}_{s+1}: y\in \mathcal{V}\rightarrow f_{s+1}(y)$. For
$i\in \{1, \ldots, s\}$, we denote by $D_i$ the degree of $f_i$ and by
$D=\max(D_i, i=1, \ldots, s+1)$.

We prove in this section that the sum of the degrees of the
equidimensional components of the critical locus of the polynomial
mapping $\widetilde{f}_{s+1}$ is bounded by $D_1\cdots D_s
(D-1)^{n-s}\binom{n}{n-s}$.

\begin{definition}\label{defdebase}
Consider an algebraic variety $\mathcal{V}\subset\C^n$, and denote by
$I(\mathcal{V})\subset\Q[X_1, \ldots, X_n]$ the ideal associated to
$\mathcal{V}$.
\begin{itemize}
  \item If $f$ is a polynomial in $\Q[X_1, \ldots, X_n]$, the {\em linear
  part} of $f$ at a point $p=(p_1, \ldots, p_n)\in\C^n$, denoted by $d_p(f)$,
  is defined to be: $d_p(f)=\frac{\partial f}{\partial X_1}(X_1-p_1)+\ldots+\frac{\partial
  f}{\partial X_n}(X_n-p_n)$. 
  \item The tangent space of $\mathcal{V}$ at $p$, denoted by
  $T_p(\mathcal{V})$, is the set of common zeroes of $d_p(f)$ for $ f\in
  I(\mathcal{V})$.
  \item For $p\in \mathcal{V}$, the dimension of $\mathcal{V}$ at $p$, denoted
  by $\dim_p(\mathcal{V})$, is the maximum dimension of an irreducible
  component of $\mathcal{V}$ containing $p$. 
  \item A point $p\in\mathcal{V}$ is said to be {\em smooth} (or {\em
  nonsingular}) if $\dim(T_p(\mathcal{V}))=\dim_p(\mathcal{V})$.
  \item An algebraic variety $\mathcal{V}\subset\C^n$ is {\em smooth} if and
  only if all points $p\in \mathcal{V}$ are smooth points.
\end{itemize}
\end{definition}

\begin{lemma}\label{lemme:1}
Let $\mathcal{V}\subset\C^n$ be a {\em smooth} algebraic variety
defined by $s$ polynomials $f_1, \ldots, f_s$ in $\Q[X_1, \ldots,
X_n]$. Suppose $\<f_1, \ldots, f_s\>$ to be radical, and let $f_{s+1}$
be a polynomial in $\Q[X_1, \ldots, X_n]$ and $\widetilde{f}_{s+1}$ be the
mapping:
$$
\begin{array}{cccc}
\widetilde{f}_{s+1} : & \mathcal{V}\subset\C^n &\longrightarrow & \C
   \\ & (x_1, \ldots, x_n)&\mapsto & f_{s+1}(x_1, \ldots, x_n) \\
\end{array}
$$ Given $p\in\mathcal{V}$, the point $p$ is a critical point of
$\widetilde{f}_{s+1}$ if and only if there exists a point $(\lambda_1,
\ldots, \lambda_{s})$ in $\C^{s}$ such that $(\lambda_1, \ldots,
\lambda_{s}, p)\in \C^{s}\times\C^n$ is a solution of the polynomial
system in $\Q[\ell_1, \ldots, \ell_{s}, X_1, \ldots, X_n]$:
$$
\left\{
\begin{array}{l}
f_1=\cdots=f_s=0 \\
\medskip
\ell_1\frac{\partial f_{1}}{\partial X_1}+\cdots+\ell_{s}\frac{\partial f_{s}}{\partial X_{1}}= \frac{\partial f_{s+1}}{\partial X_{1}}\\
\medskip
\ell_1\frac{\partial f_{1}}{\partial X_2}+\cdots+\ell_{s}\frac{\partial f_{s}}{\partial X_{2}}= \frac{\partial f_{s+1}}{\partial X_{2}}\\
~~~~~~~~~~\vdots \\
\ell_1\frac{\partial f_{1}}{\partial X_n}+\cdots+\ell_{s}\frac{\partial f_{s}}{\partial X_{n}}= \frac{\partial f_{s+1}}{\partial X_{n}}\\
\end{array}
\right .
$$ where $\ell_1, \ldots, \ell_{s}$, are new variables. 
\end{lemma}

\begin{demo} By definition, a point $p\in\mathcal{V}$ is a critical point of
$\widetilde{f}_{s+1}$ restricted to $\mathcal{V}$ if and only if the
differential of $\widetilde{f}_{s+1}$ at $p$, denoted by
$d_p(\widetilde{f}_{s+1})$ is not surjective. This is equivalent to
say that the gradient $\mathbf{grad}_p(f_{s+1})$ is orthogonal to
$T_p(\mathcal{V})$. On the other hand, from the second item of
Definition~\ref{defdebase} and since $\langle f_1, \ldots, f_s\rangle$
is radical, the vector space ${\rm Span}(\mathbf{grad}_p(f_1), \ldots,
\mathbf{grad}_p(f_s))$ is supplementar with $T_p(\mathcal{V})$.

Thus, $p$ is a critical point of $\widetilde{f}_{s+1}$ restricted to $\mathcal{V}$ if and only
if the gradient $\mathbf{grad}_p(f_{s+1})$ belongs to ${\rm
Span}(\mathbf{grad}_p(f_1), \ldots, \mathbf{grad}_p(f_s))$. 


In other words, there exist complex numbers $\lambda_{1}, \ldots,
\lambda_{s}$ such that:
$$
\begin{array}{l}
f_1(p)=\cdots=f_s(p)=0 \\
\medskip
\lambda_1\frac{\partial f_{1}}{\partial X_1}+\cdots+\lambda_{s}\frac{\partial f_{s}}{\partial X_{1}}= \frac{\partial f_{s+1}}{\partial X_{1}}\\
\medskip
\lambda_1\frac{\partial f_{1}}{\partial X_2}+\ldots+\lambda_{s}\frac{\partial f_{s}}{\partial X_{2}}= \frac{\partial f_{s+1}}{\partial X_{2}}\\
~~~~~~~~~~\vdots \\
\lambda_1\frac{\partial f_{1}}{\partial X_n}+\cdots+\lambda_{s}\frac{\partial f_{s}}{\partial X_{n}}= \frac{\partial f_{s+1}}{\partial X_{n}}\\
\end{array}
$$
which ends the proof. 
\end{demo}

\begin{remark} This algebraic characterization is well known as {\em
Lagrange's characterization} (or {\em Lagrange's system}).

\medskip
Note that the above Lemma defines critical points of a polynomial
mapping restricted to an algebraic variety as roots of an elimination
ideal. Remark that the considered algebraic variety is not supposed to
be equidimensional contrarily to the algebraic characterization of
critical points which is used
in~\cite{aubry00b,safey_el_din01a,sasc03,BGHM3,BGHM2,BGHM1}.

This is a key point to generalize Safey-Schost's
algorithm~\cite{sasc03} computing at least one point in each connected
component of a real algebraic variety to the non equidimensional case.
\end{remark}

We are now ready to state the main result of this section. 


\begin{theorem}\label{thm:critique}
Let $f_1,\ldots,f_{s}$ in $\Q[X_1, \ldots, X_n]$ be $s$ polynomials
(with $s \leqslant n-1$). Suppose $\<f_1, \ldots, f_s\>$ is a radical
ideal and defines a smooth algebraic variety $\mathcal{V}\subset
\C^n$. Let $f_{s+1}\in\Q[X_1, \ldots, X_n]$ and consider the mapping
$\widetilde{f}_{s+1}$ sending $x\in \C^n$ to $f_{s+1}(x)$. Denote by
$D_1, \ldots D_s, D_{s+1}$ the respective degrees of $f_1, \ldots,
f_s, f_{s+1}$, and by $D$ the maximum of $D_1, \ldots, D_s,
D_{s+1}$. Then, the sum of the degrees of the equidimensional
components of the critical locus of $\widetilde{f}_{s+1}$ restricted
to $\mathcal{V}$ is bounded by:
$$
D_1\cdots D_s (D-1)^{n-s}{\binom{n}{n-s}}
$$
\end{theorem}

\begin{demo} Let $X_0$ and $\ell_0$ be new variables and
  denote,
  as in the proof of Theorem \ref{thm:affine},
  by
$\phi$ the mapping which associates to $f\in \Q[X_1, \ldots, X_n, \ell_1, \ldots,
\ell_k]$ the polynomial
$$\phi(f)=X_0^{\deg_X(f)}\ell_0^{\deg_\ell(f)}f(\frac{X_1}{X_0}, \ldots,
\frac{X_n}{X_0}, \frac{\ell_1}{\ell_0},\ldots, \frac{\ell_k}{\ell_0})$$ where
$\deg_X(f)$ (resp. $\deg_\ell(f)$) denotes the degree of $f$ when it is seen
as a polynomial in $\Q(\ell_1, \ldots, \ell_k)[X_1, \ldots, X_n]$
(resp. $\Q(X_1, \ldots, X_n)[\ell_1, \ldots, \ell_k]$). 

Denote by $J$ the ideal generated by Lagrange's system $S$ given in
Lemma~\ref{lemme:1}. Bounding the sum of the geometric degrees of the
equidimensional components of the critical locus of
$\widetilde{f}_{s+1}$ is equivalent to bounding the sum of the degrees
of the equidimensional components of $\sqrt{J}$. {F}rom
Corollary~\ref{cor:final}, this sum is bounded by the strong bi-degree
of $\sqrt{\phi(J)}$. Now, applying Theorem~\ref{bezout} to
$\langle\phi(f), f \in S\rangle$ ends the proof.
\end{demo}

\begin{remark}
Suppose $\mathcal{V}\subset\C^n$ is an equidimensional algebraic
variety of dimension $d$. In this case, the critical points of a
mapping can be characterized by the vanishing of some minors of a
jacobian matrix. Then, applying the classical B\'ezout theorem to the
obtained polynomial system yields the degree bound:
$$D^{n-d}\left ( (n-d)(D-1)\right )^d.$$

This quantity, which is greater than the one obtained in
Theorem~\ref{thm:critique}, is used in~\cite{BGHM1,BGHM2} to bound, in
the worst case, the number of critical points computed by the
algorithms proposed in these papers. A similar approach is used
in~\cite{safey_el_din01a,BGHM3,BGHM4}. 

Our bound, given in Theorem~\ref{thm:critique}, shows that the
previous ones were not sharp, in particular when $d=n/2$.
\end{remark}

The following corollary is intensively used in the next section. 

\begin{corollary}\label{corol:bound}
Let $f_1,\ldots,f_{s}$ in $\Q[X_1, \ldots, X_n]$ be $s$ polynomials
(with $s \leqslant n-1$). Suppose they generate a radical ideal and
define a smooth algebraic variety $\mathcal{V}\subset \C^n$. Denote by
$D_1, \ldots D_s$ the respective degrees of $f_1, \ldots, f_s$, and by
$D$ the maximum of $D_1, \ldots, D_s$.  Consider $\pi:
\C^n\rightarrow\C$ the canonical projection on the first coordinate
and suppose the critical locus of its restriction to $\mathcal{V}$ to
be zero-dimensional. Then, the degree of the critical locus of $\pi$
restricted to $\mathcal{V}$ is bounded by:
$$
D_1\cdots D_s (D-1)^{n-s}{\binom{n}{n-s}}
$$

\medskip
Moreover, if $(f_1, \ldots, f_s)$ is a regular sequence, the degree of the critical
locus of $\pi$ restricted to $\mathcal{V}$ is bounded by:
$$
D_1\cdots D_s (D-1)^{n-s}{\binom{n-1}{n-s}}
$$
\end{corollary}

\begin{demo} The first item is a direct application of 
Theorem~\ref{thm:critique}.

We focus on the case where $(f_1, \ldots, f_s)$ is a regular
sequence. {F}rom Lemma~\ref{lemme:1}, the critical locus of the
restriction of $\pi$ to $\mathcal{V}$ is the projection on $X_1,
\ldots, X_n$ of the zero-set of
\begin{equation}\label{syst1}
\left\{
\begin{array}{l}
f_1=\cdots=f_s=0,  \\
\medskip
\medskip
\ell_1\frac{\partial f_1}{\partial X_1}+\cdots+\ell_{s}\frac{\partial f_{s}}{\partial X_{1}}= 1\\
\ell_1\frac{\partial f_1}{\partial X_2}+\cdots+\ell_{s}\frac{\partial f_{s}}{\partial X_{2}}= 0\\
~~~~~~~~~~~~~\vdots \\
\ell_1\frac{\partial f_1}{\partial X_n}+\cdots+\ell_{s}\frac{\partial f_{s}}{\partial X_{n}}= 0\\
\end{array}
\right .
\end{equation}
Since the critical locus of the restricition of $\pi$ to $\mathcal{V}$
is supposed to be zero-dimensional, there exists a Zariski-closed
subset $\mathcal{A}\subset \C^{s-1}$ such that chosing $(a_1, \ldots,
a_{s-1})\subset \C^{s-1}\setminus \mathcal{A}$ and substituting $f_s$
by $f_s+a_1 f_1+\cdots+a_{s-1} f_{s-1}$ in (\ref{syst1}) yields a
polynomial system such that for any of its solution $(x_1, \ldots,
x_n, \lambda_1, \ldots, \penalty-10000\lambda_s)$, $\lambda_s\neq 0$.
Thus, one can suppose that all solutions of (\ref{syst1}) satisfy
$\ell_s\neq 0$.


Then, the set of critical points of $\pi$ restricted to $\mathcal{V}$ is
contained in the projection on $X_1, \ldots, X_n$ of the zero-set of: 
\begin{equation}\label{syst2}
\left\{
\begin{array}{l}
f_1=\cdots=f_s=0 \\
\medskip
\medskip
m_1\frac{\partial f_1}{\partial X_2}+\cdots+m_{s-1}\frac{\partial f_{s-1}}{\partial X_{2}}+\frac{\partial f_{s}}{\partial X_{2}}= 0\\
~~~~~~~~~~\vdots \\
m_1\frac{\partial f_1}{\partial X_n}+\cdots+m_{s-1}\frac{\partial f_{s-1}}{\partial X_{n}}+\frac{\partial f_{s}}{\partial X_{n}}= 0\\
\end{array}
\right .
\end{equation}
We prove now the reverse inclusion. 


Consider a solution $p=(x_1, \ldots, x_n, \mu_1, \ldots, \mu_{s-1})$
of (\ref{syst2}). Since $(f_1, \ldots, f_s)$ is regular sequence
defining a radical ideal and $\mathcal{V}$ is smooth, the jacobian
matrix ${\rm Jac}(f_1, \ldots, f_s)$ has full rank at $(x_1, \ldots,
x_n)$. This implies the polynomial $m_1\frac{\partial f_1}{\partial
X_1}+\cdots+m_{s-1}\frac{\partial f}{\partial X_1}+\frac{\partial f_{s}}{\partial X_{1}}$ to be not null
at $p$.  

Thus, for any solution $p=(x_1, \ldots, x_n, \mu_1, \ldots,
\mu_{s-1})$ there exists $\mu\neq 0$ such that $(x_1, \ldots, x_n,
\mu_1, \ldots, \mu_{s-1}, \mu)$ is a solution of
\begin{equation}\label{syst3}
\left\{
\begin{array}{l}
f_1=\cdots=f_s=0 \\
\medskip
\medskip
m_1\frac{\partial f_1}{\partial X_2}+\cdots+m_{s-1}\frac{\partial f_{s-1}}{\partial X_{2}}+\frac{\partial f_{s}}{\partial X_{2}}= m_s\\
m_1\frac{\partial f_1}{\partial X_2}+\cdots+m_{s-1}\frac{\partial f_{s-1}}{\partial X_{2}}+\frac{\partial f_{s}}{\partial X_{2}}= 0\\
~~~~~~~~~~\vdots \\
m_1\frac{\partial f_1}{\partial X_n}+\cdots+m_{s-1}\frac{\partial f_{s-1}}{\partial X_{n}}+\frac{\partial f_{s}}{\partial X_{n}}= 0\\
\end{array}
\right .
\end{equation}
Conversly, since ${\rm Jac}(f_1, \ldots, f_s)$ has full rank at any
point of $\mathcal{V}$, for any solution $(x_1, \ldots, x_n, \mu_1,
\ldots, \mu_s)$ of (\ref{syst3}) $\mu_s\neq 0$. Then, one can divide
each equation in (\ref{syst3}) by $m_s$ and put $\ell_1=m_1/m_s,
\ldots, \ell_{s-1}=m_{s-1}/m_s$ and $\ell_s=1/m_s$ to recover
(\ref{syst1}). This allows us to conclude that the projection of the
zero-set of (\ref{syst2}) is the critical locus of $\pi$ restricted to
$\mathcal{V}$. 

Now, applying Corollary~\ref{cor:final} and Theorem~\ref{bezout} to the system
obtained by bihomogeneizing the system (\ref{syst2}) ends the proof. \hfill

\end{demo}

We show in the following section how our bound can be used to improve the
already known bounds on the first Betti number of a smooth real algebraic variety
defined by a polynomial system generating a radical ideal.

\section{Generalization of Safey/Schost's Algorithm}\label{sec:sasc}

In this section, we first generalize to the non equidimensional case,
the algorithm provided in~\cite{sasc03} computing at least one point
in each connected component of the real counterpart of a smooth
equidimensional algebraic variety. Then, we estimate the number of
points computed by the algorithm we propose using
Theorem~\ref{thm:critique} to bound the first Betti number of a smooth
real algebraic set defined by a polynomial system generating a radical
ideal.

Given a smooth algebraic variety $\mathcal{V}\subset\C^n$ of dimension $d$, we
denote by $\Pi_i$ (for $i$ in $\{1, \ldots, d\}$) the canonical projection:
$$
\begin{array}{cccc}
\Pi_i : & \C^n &\longrightarrow &\C^i \\
    & (x_1, \ldots, x_n) & \mapsto & (x_1, \ldots, x_i) \\
\end{array}
$$ and by $W_{n-(i-1)}(\mathcal{V})$ the critical locus of the
restriction of $\Pi_{i}$ to $\mathcal{V}$, i.e. {\bf the union of the
critical points of the restrictions of $\Pi_i$ to each equidimensional
component of $\mathcal{V}$}. Following~\cite{sasc03}, we set
$W_{n-d}(\mathcal{V})=\mathcal{V}$ and we have:
$$
W_{n}(\mathcal{V})\subset W_{n-1}(\mathcal{V})\subset \ldots\subset W_{n-d+1}(\mathcal{V})\subset W_{n-d}(\mathcal{V})
$$ In the equidimensional case, the algorithm provided in
\cite{sasc03} is based on the following geometric result:

\begin{theorem}\label{thm:sasc03} \cite{sasc03}
Let $\mathcal{V}\subset\C^n$ be a smooth equidimensional algebraic variety of
dimension $d$. Up to a generic linear change of variables, given an {\rm arbitrary}
point $p=(p_1, \ldots, p_d)\in \R^d$, $W_{n-(i-1)}(\mathcal{V})\cap\Pi_{i-1}^{-1}(p_1,
\ldots, p_{i-1})$ is zero-dimensional for $i$ in $\{1, \ldots, d+1\}$, and the
union of the finite algebraic sets:
$$
W_{n-d}(\mathcal{V})\cap\Pi_{d}^{-1}(p_1, \ldots, p_d), \ldots,
W_{n-(i-1)}(\mathcal{V})\cap\Pi_{i-1}^{-1}(p_1, \ldots, p_{i-1}),\ldots, W_n(\mathcal{V})
$$
intersects each connected component of $\mathcal{V}\cap \R^n$. 
\end{theorem}

A naive way of using this result in non equidimensional situations is
to compute an equidimensional decomposition of the ideal $\langle f_1,
\ldots, f_s\rangle$ and to apply Theorem~\ref{thm:sasc03} to each
computed equidimensional component. This technique is underlying in
many recent algorithms computing at least one point in each connected
component of a real algebraic set
(see~\cite{aubry00b,sasc02,safey_el_din01a}) and does not allow us to
prove satisfactory complexity results neither on the output of the
algorithms nor on the arithmetic complexity, since the degree of the
polynomials defining each equidimensional component is not well
controlled.


  \begin{lemma}\label{lemma:gen}
  Let $(f_1, \ldots, f_s)$ be a polynomial family in $\Q[X_1, \ldots,
  X_n]$. Suppose it generates a radical ideal of dimension $d$ and
  defines a smooth algebraic variety $\mathcal{V}\subset\C^n$. Given a
  point $(p_1, \ldots, p_d)$ in $\Q^d$, consider the polynomial system
  in $\Q[ X_1, \ldots, X_n,\ell_1, \ldots, \ell_s]$:
  \begin{center}
  \begin{equation}\label{critsyst}
  \left\{\begin{array}{l}
  f_{1}=\cdots=f_{s}=0, \\
  X_1-p_1=\cdots =X_i-p_i=0 \\
  \ell_1\frac{\partial f_{1}}{\partial X_{i+1}}+\cdots+\ell_{s}\frac{\partial f_{s}}{\partial X_{i+1}}= 1\\
  \medskip
  \ell_1\frac{\partial f_{1}}{\partial X_{i+2}}+\cdots+\ell_{s}\frac{\partial f_{s}}{\partial X_{i+2}}= 0\\
~~~~~~~~~~~~~  \vdots \\
  \ell_1\frac{\partial f_{1}}{\partial X_n}+\cdots+\ell_{s}\frac{\partial f_{s}}{\partial X_{n}}= 0\\
  \end{array}\right .
  \end{equation}    
  \end{center}
  The projection of its complex solution set on $X_1, \ldots, X_n$ is
  $\Pi_i^{-1}(p_1, \ldots, p_i)\cap W_{n-i}(\mathcal{V})$.
  \end{lemma}

\begin{demo} 
Denote by $\mathcal{W}$ the projection of the complex solution set of
(\ref{critsyst}) on $X_1, \ldots, X_n$. We first prove it contains
$\Pi_i^{-1}(p_1, \ldots, p_i)\cap W_{n-i}(\mathcal{V})$, then we prove the
reverse inclusion.

Consider $(g_1, \ldots, g_k)\in\Q[X_1, \ldots, X_n]$ a polynomial family
generating a radical ideal whose associated algebraic variety,
denoted by $C_q$,
is an equidimensional component of $\mathcal{V}\subset\C^n$ of
dimension $q\leqslant d$. Let $y$ be a point in $\Pi_i^{-1}(p_1,
\ldots,p_i)\cap W_{n-i}(C_q)$. We first prove that it belongs to
$\mathcal{W}$.

Let $\mathbf{e}_1, \ldots, \mathbf{e}_{i+1}$ be the gradient vectors
of $X_1, \ldots, X_{i+1}$. Since $y\in W_{n-i}(C_q)$~:
$$
\dim({\rm Span}(\mathbf{e}_1, \ldots, \mathbf{e}_{i+1})+{\rm
Span}(\mathbf{grad}_y(g_1), \ldots, \mathbf{grad}_y(g_k))) \leqslant n-q+i. 
$$
and this implies that: 
$$
\mathbf{e}_{i+1}\in {\rm Span}(\mathbf{grad}_y(g_{1}), \ldots,
\mathbf{grad}_y(g_k),\mathbf{e}_1, \ldots, \mathbf{e}_{i}). 
$$
Since $\langle f_1, \ldots, f_s\rangle$ is radical, 
$${\rm
Span}(\mathbf{grad}_y(g_1),\ldots, \mathbf{grad}_y(g_k))={\rm
Span}(\mathbf{grad}_y(f_1),\ldots, \mathbf{grad}_y(f_s))$$
which implies that:
$$
\mathbf{e}_{i+1}\in {\rm Span}(\mathbf{grad}_y(f_{1}), \ldots,
\mathbf{grad}_y(f_s),\mathbf{e}_1, \ldots, \mathbf{e}_{i}). 
$$ Hence, there exists $(\lambda_1, \ldots, \lambda_s)\in \C^s$ such
that $(y, \lambda)\in \C^n\times \C^s$ belongs to the solution set of
(\ref{critsyst}).

Consider now $y\in \mathcal{V}$ such that there exists $\lambda\in
\C^s$ for which $(y, \lambda)$ is a solution of the
system~(\ref{critsyst}). We prove in the sequel that there exists an
equidimensional component $C$ of $\mathcal{V}$ such that $y$ belongs
to $W_{n-i}(C)$.

Since $y\in\mathcal{V}$ there exists an equidimensional component $C$ of
$\mathcal{V}$ such that $y\in C$ and let $q$ be the dimension of $C$. We prove
in the following that $y\in W_{n-i}(C)$ which is sufficient to conclude since
$y$ already belongs to $\Pi_i^{-1}(p_1, \ldots, p_i)$. Consider a set of
generators $g_1, \ldots, g_k$ of the ideal associated to $C$ and remark that
since $\langle f_1, \ldots, f_s\rangle$ is radical
$${\rm Span}(\mathbf{grad}_y(g_1),\ldots, \mathbf{grad}_y(g_k))={\rm
Span}(\mathbf{grad}_y(f_1),\ldots, \mathbf{grad}_y(f_s)).$$ Note also that
since $y\in \Pi_i^{-1}(p_1, \ldots, p_i)$, the vector-space $${\rm
Span}(\mathbf{grad}_y(g_1),\ldots, \mathbf{grad}_y(g_k), \mathbf{e}_1, \ldots,
\mathbf{e}_i)$$ is the co-tangent space of $C\cap\Pi_i^{-1}(p_1, \ldots, p_i)$
which has dimension at most $n-q+i$. Since there exists $\lambda\in \C^s$ such
that $(y, \lambda)\in \mathcal{W}$
$$
\mathbf{e}_{i+1}\in {\rm Span}(\mathbf{grad}_y(g_1),\ldots, \mathbf{grad}_y(g_k), \mathbf{e}_1,
\ldots, \mathbf{e}_i)
$$ 
which implies that
$$
\dim(\mathbf{grad}_y(g_1),\ldots, \mathbf{grad}_y(g_k), \mathbf{e}_1,
\ldots, \mathbf{e}_i, \mathbf{e}_{i+1})\leqslant n-q+i
$$ 
By definition, this proves $y$ belongs to $W_{n-i}(C)$. 

\end{demo}

Given a polynomial $f\in \Q[X_1, \ldots, X_n]$, and $\mathbf{A}\in GL_n(\Q)$,
we denote by $f^\mathbf{A}$ the polynomial obtained by performing the change
of variables induced by $\mathbf{A}$ on $f$.

\begin{lemma}\label{lemma:chgvar}
Let $\mathcal{V}\subset\C^n$ be a smooth algebraic variety defined by
$s$ polynomials $f_1, \ldots, f_s$ in $\Q[X_1, \ldots, X_n]$
generating a {\em radical} ideal.  Given $\mathbf{A}\in GL_n(\Q)$,
consider $I^\mathbf{A}_0\subset\Q[\ell_1, \ldots, \ell_{s}, X_1,
\ldots, X_n]$ the ideal generated by the polynomial system:
$$
\left\{
\begin{array}{l}
f^\mathbf{A}_1=\cdots=f^\mathbf{A}_s=0 \\
\medskip
\ell_1\frac{\partial f^\mathbf{A}_{1}}{\partial X_1}+\cdots+\ell_{s}\frac{\partial f^\mathbf{A}_{s}}{\partial X_{1}}= 1\\
\medskip
\ell_1\frac{\partial f^\mathbf{A}_1}{\partial X_2}+\cdots+\ell_{s}\frac{\partial f^\mathbf{A}_{s}}{\partial X_{2}}= 0\\
~~~~~~~~~~~~~~~~~\vdots \\
\ell_1\frac{\partial f^\mathbf{A}_1}{\partial X_n}+\cdots+\ell_{s}\frac{\partial f^\mathbf{A}_{s}}{\partial X_{n}}= 0\\
\end{array}
\right .
$$ There exists a proper Zariski-closed subset $\mathcal{H}\subset
GL_n(\C)$ such that if $\mathbf{A}\notin \mathcal{H}$,
$I_0^\mathbf{A}$ is radical and the elimination ideal
$I^\mathbf{A}_0\cap\Q[X_1, \ldots, X_n]$ is {\em zero-dimensional} or
equal to $\langle 1\rangle$.

Suppose additionally that $f_1, \ldots, f_s$ is a regular
sequence. Then, there exists a proper Zariski-closed subset
$\mathcal{H}^\prime\subset GL_n(\C)$ such that if
$\mathbf{A}\notin\mathcal{H}^\prime$, the 
ideal
$I^\mathbf{A}_0$ is radical
and is either {\em zero-dimensional} or equal to $\langle 1\rangle$.
\end{lemma}

\begin{demo} 
Consider a polynomials family $(g_1, \ldots , g_k)$ in $\Q[X_1,
\ldots, X_n]$ generating a radical equidimensional ideal whose associated
algebraic variety is an equidimensional component $C_d$ of
$\mathcal{V}\subset\C^n$. Let $\mathcal{P}^\mathbf{A}\subset\C^n$ be the
algebraic variety associated to $I^\mathbf{A}_0\cap\Q[X_1, \ldots, X_n]$ and
$\mathcal{P}_d^\mathbf{A}$ a subset of $\mathcal{P}^\mathbf{A}$ such that
$\mathcal{P}_d^\mathbf{A}$ is the intersection of $\mathcal{P}^\mathbf{A}$
and $C^\mathbf{A}_d$, the complex solution set of:
$$
g_1^\mathbf{A}=\cdots=g_k^\mathbf{A}=0. 
$$

{From}~\cite[Theorem~$2$]{sasc03}, there exists a proper Zariski closed subset
of $GL_n(\C)$ such that if $\mathbf{A}\notin \mathcal{H}_d$, the critical
locus of the restriction of the canonical projection $\pi$:
$$
\begin{array}{cccc}
\pi : & \C^n &\longrightarrow & \C \\
   & (x_1, \ldots, x_n)&\mapsto & x_1 \\
\end{array}
$$ to $C_d^\mathbf{A}$ is zero-dimensional or empty. Now, consider the
polynomial system in $\Q[X_1, \ldots, X_n, m_1, \ldots, m_k]$:
$$
\left\{
\begin{array}{l}
g_1^\mathbf{A}=\cdots=g_k^\mathbf{A}=0 \\
m_1\frac{\partial g^\mathbf{A}_1}{\partial X_1}+\cdots+m_k\frac{\partial
  g^\mathbf{A}_k}{\partial X_1}=1 \\
m_1\frac{\partial g^\mathbf{A}_1}{\partial X_2}+\cdots+m_k\frac{\partial
  g^\mathbf{A}_k}{\partial X_2}=0 \\
~~~~~~~~~~\vdots \\
m_1\frac{\partial g^\mathbf{A}_1}{\partial X_n}+\cdots+m_k\frac{\partial
  g^\mathbf{A}_k}{\partial X_n}=0 \\
\end{array}
\right .
$$ and $J^\mathbf{A}$ the ideal it generates.  {From}~Lemma~\ref{lemme:1}, the
critical locus of the restricition of $\pi$ to $C_d^\mathbf{A}$ is the
algebraic variety associated to $J^\mathbf{A}\cap\Q[X_1, \ldots, X_n]$, which
is consequently zero-dimensional if $\mathbf{A}\notin\mathcal{H}_d$.

Since $\mathcal{V}$ is smooth, the tangent space to $\mathcal{V}$ at
any point $p$ in $C_d\subset \mathcal{V}$ is equal to the tangent
space to $C_d$ at $p$, and since
$\langle g_1, \ldots, g_k\rangle$ is a radical ideal, the following
holds:
$$ 
{\rm Span}(\mathbf{grad}_p(g_1), \ldots,
\mathbf{grad}_p(g_k)) = 
{\rm Span}(\mathbf{grad}_p(f_1), \ldots,
\mathbf{grad}_p(f_s))
$$ Moreover, for $\mathbf{A}.p\in C_d^\mathbf{A}\subset
\mathcal{V}^\mathbf{A}$:
$$ 
\mathbf{A}.\left ({\rm Span}(\mathbf{grad}_p(g_1), \ldots,
\mathbf{grad}_p(g_k))\right ) = 
{\rm Span}(\mathbf{grad}_{\mathbf{A}.p}(g^\mathbf{A}_1), \ldots,
\mathbf{grad}_{\mathbf{A}.p}(g^\mathbf{A}_k))
$$
and 
$$ 
\mathbf{A}\left ({\rm Span}(\mathbf{grad}_p(f_1), \ldots,
\mathbf{grad}_p(f_s))\right ) = 
{\rm Span}(\mathbf{grad}_{\mathbf{A}.p}(f^\mathbf{A}_1), \ldots,
\mathbf{grad}_{\mathbf{A}.p}(f^\mathbf{A}_s))
$$

Thus, at each point $p_\mathbf{A}\in\mathcal{P}^\mathbf{A}_d$, 
$$
{\rm Span}(\mathbf{grad}_{p_\mathbf{A}}(g_1^\mathbf{A}), \ldots,
\mathbf{grad}_{p_\mathbf{A}}(g^\mathbf{A}_k)) = 
{\rm Span}(\mathbf{grad}_{p_\mathbf{A}}(f_1^\mathbf{A}), \ldots,
\mathbf{grad}_{p_\mathbf{A}}(f^\mathbf{A}_s))
$$

Hence, $\mathcal{P}^\mathbf{A}_d$ is exactly the algebraic variety associated
to $J^\mathbf{A}\cap\Q[X_1, \ldots, X_n]$, which is zero-dimensional if
$\mathbf{A}\notin \mathcal{H}_d$. Iterating the above on each equidimensional
component of $\mathcal{V}$ proves that there exists a Zariski-closed subset
$\mathcal{A}\subsetneq GL_n(\C)$ such that if $\mathbf{A}\in GL_n(\Q)\setminus
\mathcal{A}$, $I_0^\mathbf{A}\cap\Q[X_1, \ldots, X_n]$ is either
zero-dimensional or equal to $\langle 1\rangle$.

We prove now that there exists a Zariski-closed subset $\mathcal{H}'\subsetneq
GL_n(\C)$ such that for all $\mathbf{A}\in GL_n(\Q)\setminus \mathcal{H}'$,
$I_0^\mathbf{A}$ is radical. Consider the mapping
$$\phi :\begin{array}[t]{ccc} \C^n\times \C^s&\rightarrow& \C^n\\ (y,
\lambda_1, \ldots, \lambda_s)&\rightarrow &\left
(\sum_{i=1}^s\lambda_i\frac{\partial f_i}{\partial X_1}(y), \ldots,
\sum_{i=1}^s\lambda_i\frac{\partial f_i}{\partial X_n}(y)\right )
\end{array}
$$ {F}rom Sard's theorem
, for each equidimensional component $C$ of
$\mathcal{V}$, the set of critical values
$K(\phi, C)$ of the
restriction of $\phi$ to $C\times \C^s$ is Zariski-closed in $\C^n$. Let
$\mathcal{B}$ be the union of the sets $K(\phi, C)$ for all equidimensional
components $C$ of $\mathcal{V}$ and $(a_1, \ldots, a_n)\in \Q^n\setminus
\mathcal{A}$. We prove now that the ideal $J$ generated by:
\begin{equation}\label{systtoto}
\left\{
\begin{array}{l}
f_1=\cdots=f_s=0 \\
\medskip
\ell_1\frac{\partial f_{1}}{\partial X_1}+\cdots+\ell_{s}\frac{\partial f_{s}}{\partial X_{1}}= a_1\\
~~~~~~~~~~~~~~\vdots \\
\ell_1\frac{\partial f_1}{\partial X_n}+\cdots+\ell_{s}\frac{\partial f_{s}}{\partial X_{n}}= a_n\\
\end{array}
\right .
\end{equation} is radical. Remark first that from the above paragraph, there exists a
Zariski-closed subset $\mathcal{}$ such that if $(a_1, \ldots, a_n)\in
\Q^n\setminus \mathcal{C}$, $J\cap \Q[X_1, \ldots, X_n]$ is zero-dimensional.

Let $(y, \lambda)\in \C^n \times \C^s$ be a solution of the above polynomial
system, $C$ be the equidimensional component of $\mathcal{V}$
containing $y$;
let also 
$g_1, \ldots, g_k$ be a set of generators of the ideal associated to $C$ and
$d$ be the dimension of $C$. Since $\mathcal{V}$ is smooth and $\langle f_1,
\ldots, f_s\rangle$ is radical, the jacobian matrix ${\rm Jac}(f_1, \ldots,
f_s)$ has rank $n-d$ at $y$. Since $J\cap\Q[X_1, \ldots, X_n]$ is
zero-dimensional, and since ${\rm Jac}(f_1, \ldots, f_s)$ has rank $n-d$ at
$y$, the set $\Lambda_y\subset \C^s$ defined such that $\lambda\in \Lambda_y$ if and
only if $(y, \lambda)$ belongs to the complex solution set of $J$ has
dimension $d-(n-s)$.  Thus, the irreducible component of the algebraic variety
defined by $J$ containing $(y, \lambda)$ has dimension $d-(n-s)$. In order to
prove that $J$ is radical, it is sufficient to prove that the jacobian matrix
associated to the above polynomial system has rank $n+s-(d-(n-s))=2n-d$ at
$(y, \lambda)$.

Since, by definition, $\langle g_1, \ldots, g_k\rangle$ is radical and $C$ is
smooth, there exists a subset $\{g_{i_1}, \ldots, g_{i_{n-d}}\}\subset\{g_1,
\ldots, g_k\} $ such that the rank of the jacobian matrix ${\rm Jac}(g_{i_1},
\ldots, g_{i_{n-d}})$ at $y$ equals the rank of ${\rm Jac}(g_1, \ldots, g_k)$
which is $n-d$.

Since $(a_1, \ldots, a_n)$ is not a critical value of the restriction
of $\phi$ to $C\times \Lambda_y$, the rank of the jacobian matrix
associated to the
polynomial family $$(g_{i_1}, \ldots, g_{i_{n-d}},
\sum_{i=1}^s\ell_{i}\frac{\partial f_i}{\partial X_1},\ldots,
\sum_{i=1}^s\ell_{i}\frac{\partial f_i}{\partial X_n})$$ is maximal and then
is $2n-d$ at $(y,\lambda)$ which implies that the jacobian matrix associated
to $\sum_{i=1}^s\ell_{i}\frac{\partial f_i}{\partial X_1},\ldots,
\sum_{i=1}^s\ell_{i}\frac{\partial f_i}{\partial X_n})$ with respect to the
variables $\ell_1, \ldots, \ell_s$ has rank $n$ at $(y, \lambda)$. Since ${\rm
Jac}(f_1, \ldots, f_s)$ has rank $n-d$ at $y$, this implies that the rank of
the jacobian matrix associated to the polynomial system (\ref{systtoto})
equals $2n-d$ at $(y,\lambda)$ which ends to prove that there exists a
Zariski-closed subset $\mathcal{B}$ such that if $(a_1, \ldots, a_n)\in
\Q^n\setminus\mathcal{B}$, $J$ is radical.

Suppose now $f_1, \ldots, f_s$ to be a regular sequence in $\Q[X_1,
\ldots, X_n]$ generating a radical ideal. Then the ideal $\langle f_1,
\ldots, f_s\rangle$ is equidimensional of dimension $n-s$. Thus, at
any point of $\mathcal{V}$ the jacobian matrix ${\rm Jac}(f_1, \ldots,
f_s)$ has rank $s$. Consequently, at any point $p$ of the algebraic
variety associated to $I^\mathbf{A}_0\cap\Q[X_1, \ldots, X_n]$, the
jacobian matrix ${\rm Jac}(f_1^\mathbf{A}, \ldots, f_s^\mathbf{A})$
has rank $s$ since it equals $\mathbf{A}^{-1}. {\rm
Jac}(f_1^\mathbf{A}, \ldots, f_s^\mathbf{A})$. Moreover, there exists
a proper Zariski closed subset $\mathcal{H}\subsetneq GL_n(\C)$ such
that if $\mathbf{A}\in GL_n(\Q)\setminus \mathcal{H}$, the algebraic
variety $\mathcal{P}^\mathbf{A}$ associated to
$I^\mathbf{A}_0\cap\Q[X_1, \ldots, X_n]$ is zero-dimensional
. Thus,
for any point $p\in\mathcal{P}^\mathbf{A}$ there exists at most a
finite set of points $(\lambda_1, \ldots, \lambda_s)$ in $\P(\C)^s$
which is a solution of the linear system:
$$
\lambda_1.\mathbf{grad}_p(f_1^\mathbf{A})+\ldots+\lambda_s.\mathbf{grad}_p(f_s^\mathbf{A})=
\mathbf{u}
$$
(where all, but the first, coordinates of $\mathbf{u}$ are zero ), which ends the
proof. 
\end{demo}

  Given a polynomial family $(f_1, \ldots, f_s)\subset \Q[X_1, \ldots,
  X_n]$, $d$ the dimension of the ideal $\<f_1, \ldots, f_s\>$,
  $\mathbf{A}\in GL_n(\Q)$, and $p=(p_1, \ldots, p_d)$ an arbitrary
  point of $\Q^d$ we denote by $I_i^{\mathbf{A}, p}$ (for $i\in\{1, \ldots,
  d-1\}$) the ideal in $\Q[X_1, \ldots, X_n, \ell_1, \ldots, \ell_k]$
  generated by:
  $$
  \left \{
  \begin{array}{l}
  f^\mathbf{A}_1=\cdots=f^\mathbf{A}_s=0, \\X_1-p_1=0, \ldots, X_i-p_i=0 \\
  \ell_1\frac{\partial f^\mathbf{A}_{1}}{\partial X_{i+1}}+\cdots+\ell_1\frac{\partial
  f^\mathbf{A}_s}{\partial X_{i+1}}=1 \\
  \ell_1\frac{\partial f^\mathbf{A}_1}{\partial X_{i+2}}+\cdots+\ell_1\frac{\partial
  f^\mathbf{A}_s}{\partial X_{i+2}}=0 \\
  ~~~~~~~~~~\vdots \\
  \ell_1\frac{\partial f^\mathbf{A}_1}{\partial X_n}+\cdots+\ell_1\frac{\partial
  f^\mathbf{A}_s}{\partial X_n}=0 \\
  \end{array}
  \right .
  $$ and by $I^{\mathbf{A}, p}_d$ the ideal in $\Q[X_1, \ldots, X_n, \ell_1,
  \ldots, \ell_k]$ generated by
  $f_1^\mathbf{A}=\cdots=f_s^\mathbf{A}=X_1-p_1=\cdots=X_d-p_d=0$.  Remember
  that $I^{\mathbf{A},p}_0$ denotes the ideal generated by the polynomial system:
  $$
  \left \{
  \begin{array}{l}
  f^\mathbf{A}_1=\cdots=f^\mathbf{A}_s=0, \\
  \ell_1\frac{\partial f^\mathbf{A}_1}{\partial X_1}+\cdots+\ell_1\frac{\partial
  f^\mathbf{A}_s}{\partial X_1}=1 \\
  \ell_1\frac{\partial f^\mathbf{A}_1}{\partial X_2}+\cdots+\ell_1\frac{\partial
  f^\mathbf{A}_s}{\partial X_2}=0 \\
  ~~~~~~~~~~\vdots \\
  \ell_1\frac{\partial f^\mathbf{A}_1}{\partial X_n}+\cdots+\ell_1\frac{\partial
  f^\mathbf{A}_s}{\partial X_n}=0 \\
  \end{array}
  \right .
  $$

  The following result allows us to generalize the algorithm provided
  in~\cite{sasc03} to the non equidimensional case. 

  \begin{theorem}\label{thm:algo}
  Let $(f_1, \ldots, f_s)\subset\Q[X_1, \ldots, X_n]$ be a polynomial
  family. Suppose it generates a radical ideal and defines a smooth algebraic
  variety $\mathcal{V}\subset \C^n$ of dimension $d$. Then, there exists
  hypersurfaces $\mathcal{H}\subset GL_n(\Q)$ and $\mathcal{P}\subsetneq
  \C^{d}$ such that if $\mathbf{A}\notin \mathcal{H}$ and $p\in \Q^d\setminus
  \mathcal{P}$, 
  \begin{itemize}
    \item the ideals $I^{\mathbf{A},p}_i$ (for all $i\in\{0, \ldots, d\}$) are
  radical;
    \item the ideals $I^{\mathbf{A},p}_i\cap\Q[X_1, \ldots, X_n]$ (for all
  $i\in\{0, \ldots, d\}$) are either zero-dimensional or equal to $\<1\>$;
    \item the set of their real roots has a non-empty intersection with each
  connected component of $\mathcal{V}\cap\R^n$.
  \end{itemize}
  \end{theorem}

  \begin{demo} 
  Since the ideal $I=\langle f_1, \ldots, f_s\rangle$ is radical and has
  dimension $d$, there exists Zariski-closed subsets $\mathcal{A}\subsetneq
  GL_n(\C)$ and $\mathcal{P}\subsetneq \C^d$ such that if $\mathbf{A}\in
  GL_n(\Q)\setminus \mathcal{A}$ and $p=(p_1, \ldots, p_d)\in \Q\setminus
  \mathcal{P}$, then the ideals $\langle f_1^\mathbf{A}, \ldots,
  f_s^\mathbf{A}, X_1-p_1, \ldots, X_i-p_i\rangle$ are radical. 

  For $i=1, \ldots, d-1$, denote by $J^{\mathbf{A},p}_i\subset\Q[X_{i+1},
  \ldots, X_n]$ the ideal obtained by substituting $X_1, \ldots, X_i$ by $p_1,
  \ldots, p_i$ in the given set of generators of $I_i^{\mathbf{A},p}$. Remark
  that $I_i^{\mathbf{A},p}$ is radical if and only if $J^{\mathbf{A}, p}_i$ is
  radical. {F}rom Lemma~\ref{lemma:chgvar}, there exists a Zariski-closed
  subset $\mathcal{A}\subsetneq GL_{n-i}(\C)$ such that if $\mathbf{A}\in
  GL_{n-i}(\Q)\setminus \mathcal{A}$, $J^{\mathbf{A}, p}_i$ is radical. This
  proves the first item.

  The second item is a direct consequence of Lemma~\ref{lemma:gen}. 

  We prove now the third item. Let $C_d$ be an equidimensional component of
  $\mathcal{V}\subset \C^n$ of dimension $d$. {From} Theorem~\ref{thm:sasc03},
  given an arbitrary point $(p_1, \ldots, p_d)\in \Q^d$, there exists a proper
  Zariski closed subset $\mathcal{H}_d\subset GL_n(\Q)$ such that if
  $\mathbf{A}\notin \mathcal{H}_d$, then the union of the sets
  $\Pi_i^{-1}(p_1, \ldots, p_i)\cap W_{n-i}(C_d^\mathbf{A})$ for $i=1, \ldots,
  d$ and $W_n^\mathbf{A}(C_d)$ has a non-empty intersection with each
  connected component of the real counterpart of $\mathcal{V}$. The conclusion
   follows by applying again Lemma~\ref{lemma:gen}.
\end{demo}


  Following the above result, after a generic choice of $\mathbf{A}\in
  GL_n(\Q)$, the elimination ideals $I^\mathbf{A}_i\cap\Q[X_1, \ldots, X_n]$
  are zero-dimensional or $\<1\>$ and encode at least one point in each
  connected component of $\mathcal{V}\cap\R^n$. To obtain new bounds on the
  first Betti number of a smooth real algebraic variety, it is sufficient to
  sum the bounds on the number of the critical points which are computed by
  applying Corollary~\ref{corol:bound} to each polynomial system defining the
  ideals $I_i^\mathbf{A}$ once the variables $X_1, \ldots, X_i$ have been
  substituted by $p_1, \ldots, p_i$. This proves the following result.

  \begin{theorem}\label{boundsasc}
  Let $(f_1, \ldots, f_s)\subset\Q[X_1, \ldots, X_n]$ (with $s\leqslant
  n-1$) generating a radical ideal and defining a smooth algebraic
  variety $\mathcal{V}\subset\C^n$ of dimension $d$. Denote by $D_1,
  \ldots, D_s$ the respective degrees of $f_1, \ldots, f_s$ and by $D$
  the maximum of $D_1, \ldots, D_s$. The number of connected
  components of $\mathcal{V}\cap\R^n$ is bounded by:
  $$
  D_1\cdots D_s\sum_{i=0}^d (D-1)^{n-s-i}{\binom{n-i}{n-i-s}}
  $$
  
  Moreover, if $(f_1, \ldots, f_s)$ is a regular sequence, the number of
  connected components of $\mathcal{V}\cap\R^n$ is bounded by:
  $$
  D_1\cdots D_s\sum_{i=0}^{n-s} (D-1)^{n-s-i}{\binom{n-1-i}{n-i-s}}
  $$
  \end{theorem}

  The worst case is the case when $D_1=\ldots=D_s=D$. In this case, it is easy
  to prove that $D^s\sum_{i=0}^{n-s} (D-1)^{n-s-i}{\binom{n-1-i}{n-i-s}}$ is
  less or equal to the Thom-Milnor bound $D.(2D-1)^{n-1}$ which is the best
  known bound on the first Betti number. Computer simulations show that
  $D^s\sum_{i=0}^d (D-1)^{n-s-i}{\binom{n-i}{n-i-s}}$ is less or equal to
  $D.(2D-1)^{n-1}$ for values $D$, $n$ and $s$ between $2$ and $200$.

  At last, remark that Theorem~\ref{thm:critique} can also be used to
  bound the output of the algorithm provided
  in~\cite{aubry00b,safey_el_din01a,BGHM3,BGHM4} by the quantity $D^s
  (D-1)^{n-s}{\binom{n}{n-s}}$. A simple application of Pascal's
  triangle formula shows that this bound is greater than the one of
  Theorem~\ref{boundsasc} in the case of a regular sequence.

\section{Algorithmic issues}

As above, consider a polynomial family $(f_1, \ldots, f_s)$ in
$\Q[X_1, \ldots, X_n]$ generating a radical ideal and defining a
smooth algebraic variety $\mathcal{V}\subset \C^n$ of dimension
$d$. Let $p=(p_1, \ldots, p_d)$ be an arbitrary point of $\C^d$. The
algorithm relying on Theorem~\ref{thm:algo} consists in choosing generically
a matrix $\mathbf{A}\in GL_n(\Q)$ and 
\begin{itemize}
\item solving the polynomial systems generating the ideals $I^\mathbf{A}_i$
(for $i=1, \ldots, d-1$):
  $$
  \left \{
  \begin{array}{l}
  f^\mathbf{A}_1=\cdots=f^\mathbf{A}_s=0, \\X_1-p_1=0, \cdots, X_i-p_i=0 \\
  \ell_1\frac{\partial f^\mathbf{A}_1}{\partial X_{i+1}}+\cdots+\ell_1\frac{\partial
  f^\mathbf{A}_s}{\partial X_{i+1}}=1 \\
  \ell_1\frac{\partial f^\mathbf{A}_1}{\partial X_{i+2}}+\cdots+\ell_1\frac{\partial
  f^\mathbf{A}_s}{\partial X_{i+2}}=0 \\
  ~~~~~~~~~~\vdots \\
  \ell_1\frac{\partial f^\mathbf{A}_1}{\partial X_n}+\cdots+\ell_1\frac{\partial
  f^\mathbf{A}_s}{\partial X_n}=0 \\
  \end{array}
  \right .
  $$ 
\item solving the polynomial system
  $$
  \left \{
  \begin{array}{l}
  f^\mathbf{A}_1=\cdots=f^\mathbf{A}_s=0, \\
  \ell_1\frac{\partial f^\mathbf{A}_1}{\partial X_{1}}+\cdots+\ell_s\frac{\partial
  f^\mathbf{A}_s}{\partial X_{1}}=1 \\
  \ell_1\frac{\partial f^\mathbf{A}_1}{\partial X_{2}}+\cdots+\ell_s\frac{\partial
  f^\mathbf{A}_s}{\partial X_{2}}=0 \\
  ~~~~~~~~~~\vdots \\
  \ell_1\frac{\partial f^\mathbf{A}_1}{\partial X_n}+\cdots+\ell_s\frac{\partial
  f^\mathbf{A}_s}{\partial X_n}=0 \\
  \end{array}
  \right .
  $$ 
generating the ideal $I_0^\mathbf{A}$;
\item and solving the polynomial system
$f_1^\mathbf{A}=\cdots=f_s^\mathbf{A}=X_1-p_1=\cdots=X_d-p_d=0$ generating the
ideal $I^\mathbf{A}_d$.
\end{itemize}

Note that by solving a zero-dimensional polynomial system in a polynomial ring
$\Q[v_1, \ldots, v_{n+s}]$, we mean computing a
rational parameterization of its solutions. 
$$\left \{
\begin{array}{l}
v_{n+s}=\frac{q_{n+s}(T)}{q_0(T)} \\
~~~~~~~\vdots \\
v_1=\frac{q_1(T)}{q_0(T)} \\
f(T)=0
\end{array}\right .
$$
where $f, q_0, q_1, \ldots, q_{n+s}$ are univariate polynomials with coefficients
in $\Q$. 

Here, the polynomial systems we want to solve generate positive dimensional
ideals $I_i^\mathbf{A}\subset \Q[X_1, \ldots, X_n, \ell_1, \ldots, \ell_s]$
whom intersections with $\Q[X_1, \ldots, X_n]$ are zero-dimensional. 

Thus, in order to retrieve the zero-sets of $I_i^\mathbf{A}\cap\Q[X_1, \ldots,
X_n]$ (for $i=0, \ldots, d$), it is enough to compute rational
parameterizations of each equidimensional component $C_p$ of dimension $p$ of
the complex solution set of $I_i^\mathbf{A}$ intersected with $p$ generic
linear forms in $\Q[X_1, \ldots, X_n, \ell_1, \ldots, \ell_k]$. This is
equivalent to
\begin{itemize}
  \item perform a generic linear change of variables $\mathbf{B}\subset
GL_{n+s}(\Q)$ sending the vector of coordinates $[X_1, \ldots, X_n, \ell_1,
\ldots, \ell_s]$ to a new vector of coordinate $[v_1, \ldots, v_{n+s}]$
  \item compute a rational parameterization of each equidimensional component
  $C_p$ of dimension $p$ of the complex solution set of $I_i^\mathbf{A}$
  intersected with the linear subspace defined by $v_1=\cdots=v_p=0$
  \item retrieve the complex solution set of $I_i^\mathbf{A}\cap\Q[X_1,
  \ldots, X_n]$ by multiplying $\mathbf{B}^{-1}$ with the vector $(q_1/q_0,
  \ldots, q_{n+s}/q_0)$ for each computed parameterization and keep only the
  first $n$ coordinates of the computed vector. 
\end{itemize}
Once this computation is performed, one obtains rational parametrizations of
at least one point in each connected component of $\mathcal{V}\cap\R^n$
expressed in the coordinates obtained after the linear change of variables
induced by $\mathbf{A}$. Retrieving their coordinates in the original system
of coordinates is done by multiplying $\mathbf{A}^{-1}$ the previously
computed parameterizations. Thus, the cost of this operation is polynomial in
$n$ and linear in the degree of each computed rational parametrization. We see
below that this cost is negligible compared to the rest of the computation.

Computing rational parameterizations of the complex solution set of a
zero-dimensional ideal can be done from a Gr\"obner basis
(see~\cite{F4,F5}) using linear algebra
methods (see~\cite{AlBeRoWo94,Rou99} and references therein). Other
methods based on the representation of polynomials by straight-line
programs are provided
in~\cite{GiHeMoPa95,GiHaHeMoMoPa97,GiHeMoMoPa98,GiLeSa01,lecerf2002}.
The arithmetic complexity of our algorithm depends on the arithmetic
complexity of the chosen routine performing algebraic elimination.

Without additional algebraic informations on the systems generating
$I^\mathbf{A}_i$ such as regularity or semi-regularity (see
\cite{BaFaSa}), Hilbert-regularity can not be satisfactorily
bounded. Thus, at the time being, when using Gr\"obner bases in the
solving process, one can not give a better upper bound than one which
is doubly exponential in the number of variables \cite{MaMe98} for our
algorithm. Investigating algebraic properties of $I^\mathbf{A}_i$
could yield better bounds and this is left to a further work in the
spirit of \cite{BaFaSa}.

In~\cite{lecerf2002}, the author follows the ideas of
\cite{GiHeMoPa95,GiHaHeMoMoPa97,GiHeMoMoPa98,GiLeSa01} and extends them to
provide a probabilistic incremental algorithm computing generic fibers of the
equidimensional components of an algebraic variety; these fibers are encoded
by geometric resolutions. 

In the sequel, $g_1^\mathbf{A}$ denotes the polynomial $\ell_1\frac{\partial
f_1^\mathbf{A}}{\partial X_1}+\cdots+\ell_s\frac{\partial
f^\mathbf{A}_s}{\partial X_1}-1$ and $g_i$ denotes $\ell_1\frac{\partial
f^\mathbf{A}_1}{\partial X_i}+\cdots+\ell_s\frac{\partial
f^\mathbf{A}_s}{\partial X_i}$ (for $i=2, \ldots, n$).  Below,
$\mathcal{O}_{\log{}}(p)$ denotes the quantity $\mathcal{O}(p(\log{p})^a)$
(for some constant $a$) and $M(p)$ denotes the cost of multiplying two
univariate polynomials of degree $p$.  The following result provides the
arithmetic complexity of this algorithm:

\begin{theorem}\label{thmlecerf}\cite{lecerf2002}
  Consider $F_1, \ldots, F_p$ polynomials in $\Q[X_1, \ldots, X_n]$ of
    degree boun\-ded by ${D}$, represented by a Straight-Line
    Program of length $\mathcal{L}$ and defining a zero-dimensional
    variety. There exists an algorithm computing a geometric
    resolution of $V(F_1, \ldots, F_p)$ whose arithmetic complexity
    is:
  $$\mathcal{O}_{\log{}}(p
    n^4(n\mathcal{L}+n^3)M({D}\mathfrak{d})^3)$$ where
    $\mathfrak{d}$ is the maximum of the sums of the algebraic degrees
    of the irreducible components of the intermediate varieties
    defined by $F_1, \ldots, F_i$ for $i$ in $1, \ldots, p$.
\end{theorem}

{F}rom Theorem~\ref{thm:algo} and Corollary~\ref{corol:bound}, the
maximum of the algebraic degrees of the irreducible components of the
intermediate varieties defined by $f^\mathbf{A}_1, \ldots,
f^\mathbf{A}_i$ (for $1\leqslant i \leqslant s$) and $f^\mathbf{A}_1,
\ldots, f^\mathbf{A}_s, g^\mathbf{A}_1, \ldots, g^\mathbf{A}_i$ (for
$1\leqslant i\leqslant n$) is bounded by
$D^s(D-1)^{n-s}{\binom{n}{n-s}}$.

The polynomial system defining $I^\mathbf{A}_0$ has $n+s$ variables
and contains $n+s$ polynomials.

Moreover, given a straight-line program of length $\mathcal{L}$ evaluating the
system $(f_1, \ldots, f_s)$, using the result of ~\cite{Baur}, one can
construct a straight-line program of length $\mathcal{O}((\mathcal{L}+n^2))$
evaluating the polynomial system defining $I^\mathbf{A}_0$.

This discussion allows us to state the following result: 

\begin{theorem}\label{thm:complexity}
Let $(f_1, \ldots, f_s)$ be a polynomial family of $\Q[X_1, \ldots,
X_n]$ generating a radical ideal and defining a smooth algebraic
variety $\mathcal{V}\subset \C^n$. Denote by $D$ be the maximal degree
of $f_i$ (for $i=1,\ldots, s$) and by $L$ be the length of a
straight-line program evaluating $(f_1, \ldots, f_s)$. There exists a
probabilistic algorithm computing at least one point in each connected
component of $\mathcal{V}\cap \R^n$ in:
$$\mathcal{O}_{\log{}}((n+s)^5((n+s)(L+n^2)+(n+s)^3)M(D\mathfrak{d})^3)$$
operations in $\Q$ where $\mathfrak{d}$ is dominated by $D^s
(D-1)^{n-s}{\binom{n}{n-s}}$.
\end{theorem}

\begin{remark}
Remark that when $(f_1, \ldots, f_s)$ is a regular sequence,
$\mathfrak{d}$ is dominated by $D^s(D-1)^{n-s}{\binom{n-1}{n-s}}$.

Moreover, the above discussion allows us to state that the algorithm
of~\cite{lecerf2002} has a satisfactory complexity in the cases
obtained by dehomogenizing a bi-homoge\-neous system since the degree
of intermediate varieties it studies is bounded by bi-homogeneeous
B\'ezout bounds from Theorem~\ref{thm:affine}.
\end{remark}

Now, we discuss how this result improves the preceding ones. First remark that
our algorithm is {\em probabilistic}: this is first due that we have to avoid
Zariski closed subsets for the choices of the matrix $\mathbf{A}$ on one hand
and the point $p$ on the other hand; and the algorithm provided in
\cite{lecerf2002} computing geometric resolutions is also probabilistic. The
algorithm provided by~\cite{BPR98} is {\em deterministic} and has a complexity
which is dominated by $(4D)^{\mathcal{O}(n)}$, when the number of operations
is counted in a {\em Puiseux series field}. We emphasize that the
zero-dimensional polynomial system studied by this algorithm has always a
degree equal to $(4D)^n$ regardless of the original structure of the studied
variety. Thus, if the random choices performed during our algorithm are
correct, our algorithm improves the one of~\cite{BPR98} since our worst case
complexity involves a degree bound which is $D^s
(D-1)^{n-s}{\binom{n}{n-s}}$. Nevertheless, remark that the algorithm
of~\cite{BPR98} stands for any case, while ours requires to deal with a
reduced polynomial family defining a smooth algebraic variety. 

The strategies developed
in~\cite{aubry00b,safey_el_din01a,BGHM3,BGHM4,sasc03}, which are still
probabailistic are only valid in the equidimensional case and relies on
characterizing the critical points by the vanishing of some minors of the
jacobian matrix of the input polynomial system. Remark that this assumption is
no more required by our algorithm.

The complexity estimations of the probabilistic algorithms provided
in~\cite{BGHM3,BGHM4} involve a combinatorial quantity ${\binom{n}{n-d}}$ and
a geometric degree which is dominated by $D^{(n-d)}((n-d)(D-1)+1)^d$ where $d$
denotes the dimension of the studied algebraic variety. These algorithms study
${\binom{n}{n-d}}$ regular sequences defining critical points. These sequences
are formed of the input polynomials and of some extracted minors of their
jacobian matrix after localization. Our work allows us to avoid the
combinatorial quantity ${\binom{n}{n-d}}$ and to bound the number of computed
critical points. Nevertheless, up to our knowledge, ensuring that the {\em
intermediate} degrees appearing in these algorithms do not exceed the
bi-homogeneous bound is not immediate and can be the subject of a further
study.

Remark also that if the polynomials defining the studied variety are
quadratic, our algorithm has a complexity which is polynomial in the number of
variables and exponential in the number of equations. In~\cite{GrPa}, the
authors provide an algorithm which is dedicated to the quadratic case having a
similar complexity and which uses also Lagrange's system to characterize
critical points. The algorithm of \cite{GrPa} deals also with singular
situations.

Finally, our algorithm generalizes the one of~\cite{sasc03} since the
equidimensional assumption is dropped. We emphasize that some computer
experiments show that the quantity $D^{(n-d)}((n-d)(D-1))^d$ bounding
the maximal degree of {\em intermediate} varieties appearing
in~\cite{sasc03} is reached in some cases, at intermediate steps of
the algorithm. This problem is solved by our contribution.

\paragraph*{Perspectives.} Obtaining an efficient implementation from 
this work is not an easy task. The algorithm designed
in~\cite{lecerf2002} performs a linear change of variables which does
not take into account the bi-homogeneous structure of Lagrange's
system. Thus, some improvements could be brought in a further study.

Using Gr\"obner bases inside our algorithm has also to be
investigated. Implementations of~\cite{F4} and~\cite{F5} are at the
time being the most efficient for algebraic elimination: the doubly
exponential behavior of Gr\"obner bases is exceptional and is
restricted to very particular polynomial systems.

Finally, dropping the assumptions of our algorithm is a work in
progress, whose aim is to provide an algorithm working in any case
with a complexity which is polynomial in $n$, $s$, the bi-homogeneous
B\'ezout bound $D^s(D-1)^{n-s}{\binom{n}{n-d}}$ and the complexity of
evaluation of the input system.

\end{document}